\documentclass[a4paper,12pt]{amsart}
\pdfoutput=1

\usepackage{amssymb}
\usepackage{amsmath}
\usepackage{amsfonts}
\usepackage{tikz}

\def\Id{\mathbf{I}}
\def\ad{\mathrm{ad}}

\def\Auth{\mathrm{Aut}}
\def\ep{\epsilon}
\def\ga{\gamma}
\def\al{\alpha}
\def\th{\theta}
\def\be{\beta}
\def\de{\delta}
\def\half{\dfrac{1}{2}}
\def\htt{\hat t}
\def\hv{\hat v}
\def\htau{\hat \tau}
\def\Fh{\hat F}
\def\Rh{\hat R}

\def\d{\partial}

\def\Hom{\mathrm{Hom}}

\newcommand{\br}[1]{\left( #1 \right) }
\newcommand{\ba}[1]{\left\langle #1 \right\rangle }
\newcommand{\bd}[1]{\left\{ #1 \right\} }
\newcommand{\su}[2]{\mathop{\sum}_{#1}^{#2} }
\newcommand{\Omo}[2]{\Omega_{#1,#2;1,0} }
\newcommand{\pdif}[2]{\dfrac{\partial #1 }{\partial #2 } }
\newcommand{\pdifd}[3]{\dfrac{\partial^2 #1 }{\partial #2 \partial #3 } }
\newcommand{\mathpic}[1]{\ensuremath{\vcenter{\hbox{\begin{tikzpicture} #1 \end{tikzpicture}}}}}

\newtheorem{theorem}{Theorem}[section]
\newtheorem{proposition}[theorem]{Proposition}

\theoremstyle{definition}

\newtheorem{notation}[theorem]{Notation}

\theoremstyle{remark}
\newtheorem{remark}[theorem]{Remark}

\title[Givental graphs and inversion symmetry]{Givental graphs and inversion symmetry}

\author{P.~Dunin-Barkowski}

\author{S.~Shadrin}

\author{L.~Spitz}



\address{P.~D.-B.: Korteweg-de~Vries Institute for Mathematics, University of Amsterdam, P.~O.~Box 94248, 1090 GE Amsterdam, The Netherlands and ITEP, Moscow, Russia}
\email{P.Dunin-Barkovskiy@uva.nl}

\address{S.~S.: Korteweg-de~Vries Institute for Mathematics, University of Amsterdam, P.~O.~Box 94248, 1090 GE Amsterdam, The Netherlands}
\email{S.Shadrin@uva.nl}

\address{L.~S.: Korteweg-de~Vries Institute for Mathematics, University of Amsterdam, P.~O.~Box 94248, 1090 GE Amsterdam, The Netherlands}
\email{L.Spitz@uva.nl}

\begin{document}

\begin{abstract} Inversion symmetry is a very non-trivial discrete symmetry of Frobenius manifolds. It was obtained by Dubrovin from one of the elementary Schlesinger transformations of a special ODE associated to a Frobenius manifold. In this paper, we review the Givental group action on Frobenius manifolds in terms of Feynman graphs and obtain an interpretation of the inversion symmetry in terms of the action of the Givental group. We also consider the implication of this interpretation of the inversion symmetry for the Schlesinger transformations and for the Hamiltonians of the associated principle hierarchy. 
\end{abstract}

\maketitle


\section{Introduction}

A Frobenius manifold is a differential-geometric structure that was introduced by Dubrovin in the early 1990's as a mathematical framework for the study of two-dimensional topological field theory in genus zero~\cite{Dub93,Dub96}. It has appeared to be a quite universal structure that has many naturally arising examples. In particular, Frobenius manifolds can serve as a classification tool for (dispersionless) bi-Hamiltonian hierarchies of hydrodynamic type~\cite{DubZha98,DubZha05}. Nowadays there is a number of standard textbooks on Frobenius manifolds, see~\cite{Dub96,Man99,Her02}.

One of the key questions is what would be a proper extension of a given structure of Frobenius manifold to higher genera. The main result states that  under some assumptions (semi-simplicity and homogeneity) there is an unambiguous genus expansion for a Frobenius manifolds. In different languages this can be described as a reconstruction of a dispersive hierarchy from its dispersionless limit, as a classification of cohomological field theories (that is, representations of the modular operad of homology of moduli spaces of curves with marked points), and as an explicit formula that emulates the localization formula for the Gromov-Witten partition functions of the projective spaces. 

In early 2000's Givental, van de Leur, and Losev-Polyubin independently observed that there exists an action of the loop group of $GL(n)$ on the space of $n$-dimensional Frobenius manifolds~\cite{Giv01,Leu01,LosPol01}, see also~\cite{CheKonSch05}. Givental proposed a quantization of this group action and, in terms of the quantized group elements, an explicit formula for the genus expansion that we mentioned above. The three different approaches to this group action are identified in~\cite{FeiLeuSha10,ShaZvo09}.

Nowadays Givental's quantized group action seems to be the most important tool in the theory of Frobenius manifolds (and, in particular, in Gromov-Witten theory) used in most of the applications. Its applications include new relations between Gromov-Witten and Fan-Jarvis-Ruan-Witten theory and integrable hierarchies, formulations of the crepant resolution conjecture, mirror symmetry, general properties of the dispersive hierarchies constructed by Dubrovin and Zhang, relations to the homotopy BV algebras and BCOV theory, and this is still an incomplete list.

One of the most convenient tools for applying of Givental's theory is the explicit computation of the Lie algebra action of the Givental group, first proposed and used by Y.-P.~Lee~\cite{Lee08,Lee09}. The computation of the infinitesimal deformations of various structures associated with Frobenius manifolds allows us to understand the general properties of these structures, but it is usually very hard to obtain explicit formulas for the action of the particular elements of the Givental group. Let us mention two examples where the Givental action can be computed explicitly and appears to represent known important constructions. One is the BCOV mirror construction the way it was explained in~\cite{Sha09}, another one is Shramchenko's family of Frobenius structures associated to a Hurwitz space~\cite{BurSha10}.

In~\cite{Dub96}, Dubrovin derived some symmetries of Frobenius manifolds coming from the elementary Schlesinger transformations of the associated special ODE. One type of transformations, the so-called Legendre-type transformations, refers to the possible choices of flat coordinates for the associated pencil of flat connections that let it be integrated to a solution of the WDVV equation (we are not sure that it is presented in that way anywhere, but implicitly it is explained in~\cite{LosPol01,LosPol03}). Another transformation is called the \emph{inversion symmetry} and it really looks completely unexpected in terms of the solution of the WDVV equation and its flat coordinates. 

Recently, Liu, Xu, and Zhang studied the action of the inversion symmetry on the integrable hierarchies associated to Frobenius manifolds~\cite{LiuXuZha10}. They described the action of the inversion symmetry on the principal (dispersionless) hierarchies completely; it turns out to be a particular reciprocal transformation. They made some interesting conjectures on the topological deformations of those hierarchies and the genus expansion of the corresponding tau-function.

\subsection{Goals of the paper} The main result of this paper is an explanation of the inversion symmetry in terms of the Givental group action. We find it very interesting for several reasons. First, we really feel it to be a theoretical gap if some discrete symmetry of Frobenius manifolds is not understood in universal Givental terms. Second, at least formally, this gives a complete description of the transformation of the associated principal hierarchy and its topological deformation since the corresponding Lie algebra action on the bracket, the Hamiltonians, and the equations are computed in~\cite{BurPosSha10,BurPosSha11}. 

Third, the Givental operator representing the inversion transformation appears to be very simple, it is just the exponent of the quantization of a particular matrix unit multiplied by $z$, and we can really integrate it in order to obtain explicit formulas for different natural pairs of Frobenius structures. Thus, it provides a new example where the infinitesimal action of the Lie algebra of the Givental group can be integrated to the group action. This way we reproduce the results of Liu, Xu, and Zhang on the Hamiltonians of the principal hierarchy under inversion symmetry as well as Dubrovin's original action on the associated differential operator (all the ingredients of that differential operator associated to a Frobenius manifold are best reproduced in the multi-component KP approach of van de Leur~\cite{Leu01,FeiLeuSha10,BurSha10}, and the corresponding formulas for the Lie algebra action of the Givental group were computed in~\cite{FeiLeuSha10,BurSha10}).

Meanwhile, the paper contains an exposition of the original Givental formalism in terms of Feynman diagrams. In some sense, there is nothing new there, though we propose this way of thinking about Givental group as the most convenient one for the practical need to find a particular element that maps one given Frobenius structure to another one. The inversion symmetry can then be considered as an important example where this approach to Givental theory works especially nice.

\subsection{Organization of the paper}

In section 2 we recall Y.-P.~Lee's formulas for the operators of the infinitesimal deformation and explain them in terms of graphs. In section 3 we use the graphical representation of the Givental group action in order to find a particular group element that performs the inversion symmetry. In section 4 we reproduce the elementary Schlesinger transformation that was the origin of the inversion symmetry (for that we heavily use the results obtained in~\cite{BurSha10} in multi-KP approach to Frobenius manifold structures). Finally, in section 5 we reproduce the formulas of Liu, Xu, and Zhang for the transformation of the Hamiltonians of the principle hierarchy under the inversion symmetry (this comes as a very special case of the general deformation formulas for the Hamiltonians obtained in~\cite{BurPosSha10}).

\subsection{Acknowledgments}
We thank B.~Dubrovin, A.~Buryak and the students from the 2012 spring semester course on topological field theories for helpful remarks. We also thank the anonymous referees for many helpful remarks, substantially improving the paper. The authors were supported by the Netherlands Organization for Scientific Research (NWO) through a VIDI grant (S.~S. and L.~S.) and free competition grant 613.001.021 (P.~D.-B.). P.~D.-B. was also partially supported by the Ministry of Education and Science of the Russian Federation, by RFBR grant 10-02-00499 and by joint grant 11-01-92612-Royal Society.

\section{Givental group action as a sum over graphs}
In this section we explain an interpretation of the Givental group action~\cite{Giv01,Giv04} on cohomological field theories as a sum over graphs. 

\subsection{Cohomological field theories and Frobenius manifolds}

Consider the space of partition functions for $n$-dimensional cohomological field theories
\begin{equation}\label{eq:typeFPS}
Z = \exp(\sum_{g \geq 0} \hbar^{g-1} \mathcal{F}_g)
\end{equation} 
in variables $t^{d,\mu}$, $d\geq 0$, $\mu=1,\dots,n$. Such a partition function is always tame; the weighted degree of any monomial~$ \hbar^g t^{d_1,\mu_1}\cdots t^{d_k,\mu_k}$ occurring with non-zero coefficient is not more than~$3g - 3 + k$, where the weight of~$\hbar$ is~$0$, and the weight of~$t^{d,\mu}$ is~$d$. There is a fixed scalar product $\eta$ on the vector space $V:=\langle e_1,\dots,e_n\rangle$ of primary fields corresponding to the indices $\mu=1,\dots,n$. Furthermore, we will denote by $e_{\bf{1}}$ the vector in $V$ that plays the role of the unit in the underlying family of Frobenius algebras. 

In this paper, we will always work in flat coordinates, that is, $\eta_{\alpha \beta} = \delta_{\alpha, n - \beta}$ and~$e_{\mathbf{1}} = e_1$. 

The information of the genus zero part of a cohomological field theory is equivalent to the information of a Frobenius manifold. That is, given a cohomological field theory with genus zero partition funcion~$\mathcal{F}_0$, we obtain the potential~$F$ of a Frobenius manifold by
$$
F(t^1, \ldots, t^n) = \mathcal{F}_0(t^{d,\mu})|_{t^{d,\mu} = 0 \ \mathrm{ for }\ d > 0}
$$
where we identify $t^\mu := t^{0,\mu}$. 

On the other hand, given a Frobenius manifold we can uniquely reconstruct the genus zero descendant part using topological recursion~(\cite{Man99}). Although the construction we describe below is for the full genus expansion of a cohomological field theory, it can be restricted to the genus zero part (with or without descendants), and thus interpreted as an action on the space of Frobenius manifolds. This is what we will do in Example~\ref{ex:graphComp} and the subsequent sections. 

\begin{notation}
Define the so-called correlators 
$$
\ba{\tau_{d_1}(\alpha_1)\tau_{d_2}(\al_2) \cdots \tau_{d_k}(\al_k)}_g 
$$
by
\begin{equation}
\mathcal{F}_g = \sum \frac{\ba{\tau_{d_1}(\alpha_1)\tau_{d_2}(\al_2) \cdots \tau_{d_k}(\al_k)}_g}{|\Auth((\alpha_i,d_i)_{i=1}^k)|} t^{d_1, \alpha_1} \cdots t^{d_k, \alpha_k} ,
\end{equation}
where $|\Auth((\alpha_i, d_i)_{i=1}^k)|$ denotes the number of automorphisms of the collection of multi-indices~$(\alpha_i,d_i)$ and where the sum is such that it includes each monomial~$t^{d_1, \alpha_1} \cdots t^{d_n, \alpha_n}$ exactly once. 
\end{notation}

\subsection{Differential operators}\label{sec:diff-oper}

Let us remind the reader of the original formulation, due to Y.-P.~Lee, of the infinitesimal Givental group action in terms of differential operators~\cite{Lee03, Lee08, Lee09}.

Consider a sequence of operators $r_l\in \Hom(V,V)$, $l\geq 1$, such that the operators with odd (resp., even) indices are symmetric (resp., skew-symmetric). Then we denote by $(r_l z^l)\hat{\ }$ the following differential operator:
\begin{align}
(r_l z^l)\hat{\ } := & -(r_l)_{\bf{1}}^\mu\frac{\d}{\d t^{l+1,\mu}}
+ \sum_{d=0}^\infty t^{d,\nu} (r_l)_\nu^\mu \frac{\d}{\d t^{d+l,\mu}}
 \label{eq:Al-quintized} \\
& +\frac{\hbar}{2} \sum_{i=0}^{l-1}(-1)^{i+1} (r_l)^{\mu,\nu}\frac{\d^2}{\d t^{i,\mu}\d
t^{l-1-i,\nu}}. \notag
\end{align}

Givental observed that the action of the operators 
$$
\hat{R} := \exp(\sum_{l=1}^{\infty}(r_lz^l)\hat{\ })
$$
on formal power series preserves tameness. The main theorem of~\cite{FabShaZvo06} states that this action preserves the property that $Z$ is the generating function of the correlators of a cohomological field theory with the target space $(V,\eta)$ (see also~\cite{Kaz07,Tel07}).

\begin{remark}
The action of the operators described above is usually referred to as the action of the \emph{upper triangular group}. There is also a \emph{lower triangular group} action, but we do not consider it in the present paper.
\end{remark}

\subsection{Expressions in terms of graphs}\label{sec:expr-graphs}

We now describe the Givental action in terms of graphs. Consider a connected graph $\gamma$ of arbitrary genus, and with leaves. To such a graph we assign some additional structure. First, we choose an orientation on each edge of the graph, in an arbitrary way (the contribution of a graph will not depend on these choices). Second, to each element of the graph (a leaf, an edge, a vertex) we associate some tensor over the vector space~$V[[z]]$ (where~$z$ is a formal variable) that also depends on $\hbar$ and~$t^{d,\mu}$ for $d \geq 0$ and $1\leq \mu \leq n$. This graph equipped with an additional structure of such a type we denote by $\check\gamma$.

\begin{notation}
By a \emph{half-edge}, we mean either an edge together with a choice of one of the two adjacent vertices, or a leaf. If we want to talk only about the first of these two, we will use \emph{half of an internal edge}. 
\end{notation}

\subsubsection{Leaves}
Leaves are decorated by one of two types of vectors. The first type corresponds to the second term of the operator~\eqref{eq:Al-quintized} and is given by
\begin{equation}
\label{eq:leaf_decor} 
\mathcal{L}:=\exp\left(\sum_{l=1}^\infty r_l z^l\right) \left(\sum_{d=0}^\infty \sum_{\mu=1}^n e_\mu t^{d,\mu} z^d \right) .
\end{equation}
The second type of decoration is given by the vector
\begin{equation}
\label{eq:dshift_decor} 
\mathcal{L}_0 := -z \cdot \left(\exp(\sum_{l=1}^\infty r_l z^l) - \Id \right)(e_{\bf{1}})
\end{equation}
and corresponds to the dilaton shift (the first term of the operator~\eqref{eq:Al-quintized}). 

\subsubsection{Edges}
An edge is already oriented. We expect to decorate it with a bivector. Using the scalar product we can turn any (skew-)sym\-met\-ric operator into a bivector. However, this requires a choice of sign. Some choice of sign was already made in the differential operator~\eqref{eq:Al-quintized} when we used the symbol $(r_l)^{\mu\nu}$. Let us fix this choice. In the case of a skew-symmetric operator, the bivector is also skew-symmetric, so we have to use the orientation of the underlying edge in order to fix the ambiguity. It will be obvious later on that nothing depends on the choice of orientations on edges.

So, we are going to assign a bivector $\mathcal{E}\in (V[[z]])^{\otimes 2}$ to an oriented edge. The first copy of~$V[[z]]$ is associated to the input, the second to the output of the oriented edge. For clarity, we will denote the formal variable corresponding to the first copy by~$z$, and the one corresponding to the second copy by~$w$. We put 
$$
\mathcal{E}= \tilde{\mathcal{E}} \eta,
$$
where~$\tilde{\mathcal{E}}\in \Hom(V,V)[[z,w]]$ is given by
$$
\tilde{\mathcal{E}} := -\hbar\cdot\frac{\exp\left(\sum_{l=1}^{\infty} (-1)^{l-1}r_l z^l\right)\exp\left(\sum_{l=1}^{\infty} r_l w^l\right)-\Id}{z+w}.
$$
Let us rewrite this formula in a more convenient way. Denote by $r(z)$ the series $\sum_{l=1}^\infty r_lz^l$.
Then $\tilde{\mathcal{E}}$ is equal to
\begin{align}\label{formulaE}
\tilde{\mathcal{E}} &= -\hbar\cdot\frac{\exp (r(z)^*) \exp (r(w)) - \Id}{z+w}\\
&=-\hbar\cdot\frac{\exp \left(-r(-z)\right) \exp( r(w)) - \Id}{z+w}.\notag
\end{align}
(cf. the same formula in~\cite{Tel07}).

The change of the orientation of an edge corresponds to the replacement of an operator with its adjoint and the simultaneous interchange of $z$ and $w$. From Equation~\eqref{formulaE} it is obvious that $\tilde{\mathcal{E}}^*|_{z \leftrightarrow w}=\tilde{\mathcal{E}}$. Using the symmetry of the metric, we see that nothing depends on the choice of orientations on edges.

\subsubsection{Vertices}
The collection of correlators of order~$n$ corresponding to a formal power series~$\mathcal{F}_g$ in variables~$t^{d,\mu}$ can be considered as a tensor $\mathcal{V}_g[n]\in  (V^*[[z]])^{\otimes n}$. Namely, the tensor~$\mathcal{V}_g[n]$ sends $e_{\mu_1} z_1^{d_1} \otimes \cdots \otimes e_{\mu_n}z_n^{d_n}$ to the correlator~$\ba{\tau_{d_1}(e_{\mu_1}) \cdots \tau_{d_n}(e_{\mu_n})_g}$ (which is just a number), and we extend this definition linearly. 

We want to apply an element of the Givental group to the series $Z$; this means that we decorate the vertices of index~$n$ exactly by the tensor 
\begin{equation}
\mathcal{V}[n]:=\sum_{g\geq 0}\hbar^{g-1} \mathcal{V}_g[n].
\end{equation}

\subsubsection{Contraction of tensors}
Consider a decorated graph~$\check{\gamma}$. We  have associated vectors in $V[[z]]$ to leaves and bivectors in $(V[[z]])^{\otimes 2}$ to edges (the former depending on $\hbar$ and~$t^{d,\mu}$, the later depending on~$\hbar$). Furthermore, for each edge we have associated one copy of $V[[z]]$ with the input of the edge and the other with the output. At each vertex, we now contract the tensor~$\mathcal{V}[n]$ with the tensor product of the decorations of the half edges corresponding to the vertex, where~$n$ is the index of the vertex. The result is a number depending on~$\hbar$ and~$t^{d, \mu}$ which we denote by~$\mathcal{C}(\check{\gamma})$. 

\subsubsection{The final formula}
Finally, we sum over all possible decorated graphs like this, weighted by the inverse order of their automorphisms to obtain a formal power series in~$t^{d,\mu}$ that also depends on $\hbar$. In a formula:
\begin{equation}
\log (\hat{R} (Z)) = \sum_{\check{\gamma} \in \check{\Gamma}} \frac{1}{\# \Auth(\check{\gamma})} \mathcal{C}(\check{\gamma})
\end{equation}
where~$\check{\Gamma}$ denotes the set of all decorated graphs as above, and~$\Auth(\check{\gamma})$ is the set of automorphisms of~$\check{\gamma}$. From now on we will use a decorated graph and the function of $\hbar$ and~$t^{d,\mu}$ assigned to it by the graphical formalism interchangeably. 

It follows directly from the combinatorics of graphs that the result is represented as a formal power series of the same form as in Equation~\eqref{eq:typeFPS}.

\begin{remark}\label{re:linear}
Note that for any graph the only choice in the decoration is that for each leaf, it can either be decorated by $\mathcal{L}$ or~$\mathcal{L}_0$.

Furthermore, the decorations on the edges and leaves are defined as sums. Using the linearity of the functions with which we contract at the vertices, we can replace a graph with a leaf or edge decorated with a sum by a sum of graphs which only differ from the original one by replacing this sum with its individual terms. We will use this freedom in computations; thus, we will work graphs that are not elements of~$\check{\Gamma}$ as well.  
\end{remark}

\begin{remark}\emph{The formal variable~$z$.}
The contraction of tensors couples the power of the formal variable~$z$ to the first index of the variable~$t^{d,\mu}$. Thus, in the context of cohomological field theory, the power of~$z$ should be interpreted as keeping track of the power of the $\psi$-class appearing at the corresponding half-edge. 
\end{remark}

\subsubsection{The trivial example}
We discuss the trivial example of the Givental action, that is, we assume that $r_l=0$, $l=1,2,\dots$. In this case~$\mathcal{E}=0$, so the only connected graphs that give a non-trivial contribution are the graphs with one vertex and no edges. Furthermore, $\mathcal{L}_0$ is also zero, so we only need to compute
$$
\frac{1}{n!}\mathcal{V}[n](\underbrace{\mathcal{L}\otimes\cdots\otimes\mathcal{L}}_{n\ times}) 
$$
which is the $n$th homogeneous component of $\sum_{g\geq 0}\hbar^{g-1}\mathcal{F}_g$, as we can see directly from the definition of~$\mathcal{V}[n]$. Therefore, the sum over all graphs just gives us the initial series~$Z$.

\subsubsection{Dilaton equation and topological recursion relation}
We remind the reader of the well-known topological recursion relation and dilaton equation \cite{Wit91} which hold for any cohomological field theory.

In terms of correlators, the \textit{dilaton equation} is given by
\begin{equation}\label{eq:DilatonEq}
\ba{\tau_1(1)\tau_{b_1}(\al_1)\dots\tau_{b_k}(\al_k)}_g = (2g-2+k)\ba{\tau_{b_1}(\al_1)\dots\tau_{b_k}(\al_k)}_g
\end{equation}
for any $g$.

In terms of graphical formalism, the dilaton equation has the following interpretation: whenever we are given a graph with a leaf that is marked by $e_1 z$, the dilaton equation allows us to remove that leaf entirely, at the same time multiplying the resulting graph by $(2g-2+k)$, where $k$ is the number of leaves/edges going from the corresponding vertex (the removed leaf is not counted).
 
Consider the generating function for descendant classes
\begin{equation*}
D = \exp \br{ \sum_{d,\al} t^{d,\al} \tau_d(\al)}
\end{equation*}
Then the genus zero \textit{topological recursion relation} takes the following form (for $d_1>0$):
\begin{multline} \label{eq:TRREq}
\ba{\tau_{d_1}(\al_1)\tau_{d_2}(\al_2)\tau_{d_3}(\al_3) D}_0 = \\
\sum_{\lambda,\sigma}
\ba{\tau_{d_1-1}(\al_1)\tau_0(\lambda) D}_0 \eta^{\lambda\sigma}
\ba{\tau_0(\sigma)\tau_{d_2}(\al_2)\tau_{d_3}(\al_3) D}_0
\end{multline}

The topological recursion relation has the following graphical interpretation. Whenever we are given a graph with a leaf marked by $e_i z^k$ for some~$i$ and~$k > 0$, we can remove a~$\psi$-class (lower the power of~$z$) in the following way. Pick any two other half-edges on the same vertex (vertices in graphs that have a non-zero contribution are always at least trivalent) and split the vertex into two vertices connected by an edge marked by $\sum_{\alpha,\beta} \eta^{\alpha \beta} e_\alpha \otimes e_\beta$. Put the two chosen half-edges on one vertex and the original leaf on the other, now marked by $e_i z^{k-1}$. Take the sum over all possible distributions of the other half edges of the original vertex over the two new vertices. It is easy to see that this procedure does not depend on the choice of two half-edges, and represents the topological recursion relation. In an equation (dotted lines represent either edges which connect the vertices to some other parts of the graph or just leaves):
\def\ds{3.2}
$$
\mathpic{
[sqvert/.style={circle,draw=black,fill=white,thick,inner sep=0pt,minimum size=10pt},
rvert/.style={circle,draw=black,fill=black,thick,inner sep=0pt,minimum size=10pt},
leaf/.style={circle,draw=white,fill=white,inner sep=0pt,minimum size=0.1pt},
dott/.style={circle,draw=black,fill=black,inner sep=0pt,minimum size=1pt}
]
\node (al) at ( -.8,.8) [leaf,,label=right:$e_{\mu_1}$] {};
\node (ar) at ( .8,.8) [leaf,label=left:$e_{\mu_2}$] {};
\node (di) at ( 1.5,0) [leaf,label=right:$e_\rho z^k$] {};
\node (v) at ( 0,0) [sqvert] {};
\node (nl) at ( -0.8,-0.8) [leaf,label=left:$e_{\nu_1}$] {};
\node (nr) at (0.8,-0.8) [leaf,label=right:$e_{\nu_k}$] {};
\node(al2) at (-1.3,1.3) [leaf]{};
\node(ar2) at (1.3,1.3) [leaf]{};
\node(nl2) at (-1.3,-1.3) [leaf]{};
\node(nr2) at (1.3,-1.3) [leaf]{};
\node at ( 0,-0.8) [dott] {};
\node at ( -0.2,-0.8) [dott] {};
\node at ( 0.2,-0.8) [dott] {};	
\draw [thick] (v) -- (al);
\draw [thick] (v) -- (ar);
\draw [thick] (v) -- (nl);
\draw [thick] (v) -- (di);
\draw [thick] (v) -- (nr);
\draw [dotted] (al) -- (al2);
\draw [dotted] (ar) -- (ar2);
\draw [dotted] (nl) -- (nl2);
\draw[dotted] (nr) -- (nr2);
}
= \mathop{\sum}_{\substack{I\subseteq\{1,\dots,k\}\\ \alpha,\beta}}
\mathpic{
[sqvert/.style={circle,draw=black,fill=white,thick,inner sep=0pt,minimum size=10pt},
rvert/.style={circle,draw=black,fill=black,thick,inner sep=0pt,minimum size=10pt},
leaf/.style={circle,draw=white,fill=white,inner sep=0pt,minimum size=0.1pt},
dott/.style={circle,draw=black,fill=black,inner sep=0pt,minimum size=1pt}
]
\node (al) at ( -.8,.8) [leaf,,label=right:$e_{\mu_1}$] {};
\node (ar) at ( .8,.8) [leaf,label=left:$e_{\mu_2}$] {};
\node(al2) at (-1.3,1.3) [leaf]{};
\node(ar2) at (1.3,1.3) [leaf]{};
\node (di) at (\ds+0.8,0.7) [leaf,label=above:$e_\rho z^{k-1}$] {};
\node (v) at ( 0,0) [sqvert] {};
\node (nl) at ( -0.8,-0.8) [leaf] {};
\node (nr) at (0.8,-0.8) [leaf] {};
\node(nl2) at (-1.3,-1.3) [leaf]{};
\node(nr2) at (1.3,-1.3) [leaf]{};
\node at ( 0,-0.8) [dott] {};
\node at ( -0.2,-0.8) [dott] {};
\node at ( 0.2,-0.8) [dott] {};	
\node at ( 0,-1.1) {$\underbrace{\hspace{45pt}}$};
\node at ( 0,-1.5) {$e_{\nu_i},$};
\node at ( 0,-1.9) {$\scriptstyle i \in I$};

\node (2v) at (\ds+ 0,0) [sqvert] {};
\node (2nl) at ( \ds-0.8,-0.8) [leaf] {};
\node (2nr) at (\ds+0.8,-0.8) [leaf] {};
\node(2nl2) at (\ds-1.3,-1.3) [leaf]{};
\node(2nr2) at (\ds+1.3,-1.3) [leaf]{};
\node at ( \ds+0,-0.8) [dott] {};
\node at ( \ds-0.2,-0.8) [dott] {};
\node at ( \ds+0.2,-0.8) [dott] {};	
\node at ( \ds,-1.1) {$\underbrace{\hspace{45pt}}$};
\node at ( \ds,-1.5) {$e_{\nu_i},$};
\node at ( \ds,-1.9) {$\scriptstyle i \in \{1,\dots,k\}\setminus I$};

\draw [thick] (v) -- (al);
\draw [thick] (v) -- (ar);
\draw [dotted] (al) -- (al2);
\draw [dotted] (ar) -- (ar2);
\draw [thick] (2v) -- (di);
\draw [thick] (v) -- (nl);
\draw [thick] (v) -- (nr);
\draw [dotted] (nl) -- (nl2);
\draw[dotted] (nr) -- (nr2);

\draw [thick] (2v) -- (2nl);
\draw [thick] (2v) -- (2nr);
\draw [dotted] (2nl) -- (2nl2);
\draw[dotted] (2nr) -- (2nr2);
\draw [thick] (v) -- (2v) node [above=-2.5,align=center,near start] {$e_\alpha$} node [below,align=center,midway] {$\eta^{\alpha\beta}$} node [above=-2.5,align=center,near end] {$e_\beta$} ;
} .
$$

\subsubsection{Example; inversion symmetry in two dimensions}\label{ex:graphComp}

To illustrate the graphical formalism in practice, we explicitly compute one of the terms of the two-dimensional case of \textit{the inversion transformation} defined and studied in general in Section~\ref{sec:inversion}. Let~$F_0$ be the potential of a two-dimensional Frobenius manifold given by 
\begin{equation}
F_0(t^1,t^2) = \frac{(t^1)^2 t^2}{2} + \sum_{k \geq 3} \frac{\sigma_k}{k!} (t^2)^k 
\end{equation}
for some set of numbers $\{\sigma_k | k \geq 3\}$, and let $r(z) = \sum_k r_k z^k$ be the matrix series given by
\begin{equation}\label{Anilpotent}
r:= r_1 = \left( \begin{array}{cc} 0 & 1 \\ 0 & 0 \end{array} \right) , \quad r_k = 0 \text{ for all } k > 1 .
\end{equation} 

As above, using the topological recursion relation in genus zero we can consider $F_0(t^1,t^2)$ as a restriction to the small phase space of some full descendant genus zero potential $\mathcal{F}_0\left(\{t^{1,d},t^{2,d}\}_{d=0}^\infty\right)$, identifying $t^1,t^2$ with $t^{1,0},t^{2,0}$ and setting all other variables equal to zero. Define~$\tilde{\mathcal{F}}_0$ to be the genus zero part of $\log(\exp(\widehat{r(z)})\exp(\hbar^{-1}\mathcal{F}_0))$.
We compute the coefficient~$\tilde{\sigma}_5$ of~$(t^{2,0})^5$ in~$\tilde{\mathcal{F}}_0$ using the graphical formalism (as usual, we regard $\tilde{\mathcal{F}}_0$ as the exponential generating series for its coefficients). 

Since the variables~$t^{d,\mu}$ only appear in the formalism when we have leaves decorated by~$\mathcal{L}$, graphs contributing to~$\tilde{\sigma}_5$ must have precisely five leaves decorated by~$\mathcal{L}$.  Furthermore, in these decorations, only the terms which depend solely on~$t^{0,2}$ out of all~$t^{d,\mu}$ contribute to~$\tilde{\sigma}_5$. By equation~\eqref{Anilpotent} we have 
\begin{equation}
\exp\left(\sum_{l=1}^\infty r_l z^l\right) = 1 + r z ,
\end{equation}
therefore, after total expansion using the linearity of Remark~\ref{re:linear}, leaves that were originally decorated by~$\mathcal{L}$ have at most one~$\psi$-class. 

\begin{figure}
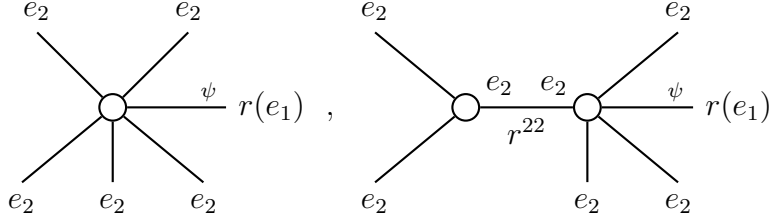

$\mathpic{
[sqvert/.style={circle,draw=black,fill=white,thick,inner sep=0pt,minimum size=10pt},
rvert/.style={circle,draw=black,fill=black,thick,inner sep=0pt,minimum size=10pt},
leaf/.style={circle,draw=white,fill=white,inner sep=0pt,minimum size=0.1pt},
dott/.style={circle,draw=black,fill=black,inner sep=0pt,minimum size=1pt}
]
\node (al) at ( -1,1) [leaf,label=above:$e_2$] {};
\node (ar) at ( 1,1) [leaf,label=above:$e_2$] {};
\node (di) at ( 1.5,0) [leaf,label=right:$r(e_1)$,label={[label distance=-2pt]100:$\scriptstyle \psi$}] {};
\node (v) at ( 0,0) [sqvert] {};
\node (nl) at ( -1.2,-1) [leaf,label=below:$e_2$] {};
\node (nc) at ( 0,-1) [leaf,label=below:$e_2$] {};
\node (nr) at (1.2,-1) [leaf,label=below:$e_2$] {};
\draw [thick] (v) -- (al);
\draw [thick] (v) -- (ar);
\draw [thick] (v) -- (nl);
\draw [thick] (v) -- (di);
\draw [thick] (v) -- (nr);
\draw [thick] (v) -- (nc);
}
\ , \ 
\mathpic{
[sqvert/.style={rectangle,draw=black,fill=white,thick,inner sep=0pt,minimum size=10pt},
rvert/.style={circle,draw=black,fill=white,thick,inner sep=0pt,minimum size=10pt},
leaf/.style={circle,draw=white,fill=white,inner sep=0pt,minimum size=0.1pt},
dott/.style={circle,draw=black,fill=black,inner sep=0pt,minimum size=1pt}
]
\node (au) at ( -2,1) [leaf,label=above:$e_2$] {};
\node (bu) at ( 2,1) [leaf,label=above:$e_2$] {};
\node (v1) at ( -0.8,0) [rvert,label={[label distance=-2pt]10:$e_2$}] {};
\node (v2) at ( 0.8,0) [rvert,label={[label distance=-2pt]170:$e_2$}] {};
\node (n2r) at (0.8,-1) [leaf,label=below:$e_2$] {};
\node (n2c) at (2.2,0) [leaf,label=right:$r(e_1)$,label={[label distance=-2pt]100:$\scriptstyle\psi$}] {};
\node (ad) at ( -2,-1) [leaf,label=below:$e_2$] {};
\node (bd) at (2,-1) [leaf,label=below:$e_2$] {};
\draw [thick] (v1) -- (v2) node [below,align=center,midway] {$r^{22}$};
\draw [thick] (v1) -- (au);
\draw [thick] (v1) -- (ad);
\draw [thick] (v2) -- (bu);
\draw [thick] (v2) -- (bd);
\draw [thick] (v2) -- (n2r);
\draw [thick] (v2) -- (n2c);
}
$
\caption{Two of the graphs contributing to~$\tilde{\sigma}_5$ where one of the leaves is decorated using~$\mathcal{L}_0$. Note that in this case, both of their contributions are zero, because $\mathcal{L}_0 = 0$. }
\label{figureDilation}
\end{figure}

In principle, there could be extra leaves which are decorated by~$\mathcal{L}_0$ (note that the variables~$t^{d,\mu}$ do not appear in~$\mathcal{L}_0$). We have drawn two graphs with such leaves in Figure~\ref{figureDilation}. However, it follows immediately from equation~\eqref{Anilpotent} that~$\mathcal{L}_0 = 0$, so the dilaton term plays no role in this computation. 

Once again using equation~\eqref{Anilpotent}, we see that in this case the edge decoration simplifies to
\begin{equation}\label{eq:E}
\mathcal{E} = - \sum_{\mu,\nu} r^{\mu,\nu} e_\mu \otimes e_\nu .
\end{equation}

By the tameness property, any vertex for which the total number of $\psi$-classes (that is, the total power of~$z$) at half-edges connected to it is equal to some~$d$, must have valence at least~$d+3$ for the graph to have a non-zero contribution. Taking into account that vertices at which no $\psi$-class appears must have either precisely three leaves, two of which are decorated with~$e_1$ and one with~$e_2$, or only leaves decorated with~$e_2$, it is easy to see that~$\tilde{\sigma}_5$ is given by the following sum: 
{\allowdisplaybreaks
\begin{multline*}
\frac{1}{5!} \tilde{\sigma}_5 = \frac{1}{5!}
\mathpic{
[sqvert/.style={circle,draw=black,fill=white,thick,inner sep=0pt,minimum size=10pt},
rvert/.style={circle,draw=black,fill=black,thick,inner sep=0pt,minimum size=10pt},
leaf/.style={circle,draw=white,fill=white,inner sep=0pt,minimum size=0.1pt},
dott/.style={circle,draw=black,fill=black,inner sep=0pt,minimum size=1pt}
]
\node (al) at ( -1,1) [leaf,label=above:$e_2$] {};
\node (ar) at ( 1,1) [leaf,label=above:$e_2$] {};
\node (v) at ( 0,0) [sqvert] {};
\node (nl) at ( -1.2,-1) [leaf,label=below:$e_2$] {};
\node (nc) at ( 0,-1) [leaf,label=below:$e_2$] {};
\node (nr) at (1.2,-1) [leaf,label=below:$e_2$] {};
\draw [thick] (v) -- (al);
\draw [thick] (v) -- (ar);
\draw [thick] (v) -- (nl);
\draw [thick] (v) -- (nr);
\draw [thick] (v) -- (nc);
}
+
\frac{1}{4!}
\mathpic{
[sqvert/.style={circle,draw=black,fill=white,thick,inner sep=0pt,minimum size=10pt},
rvert/.style={circle,draw=black,fill=black,thick,inner sep=0pt,minimum size=10pt},
leaf/.style={circle,draw=white,fill=white,inner sep=0pt,minimum size=0.1pt},
dott/.style={circle,draw=black,fill=black,inner sep=0pt,minimum size=1pt}
]
\node (al) at ( -1,1) [leaf,label=above:$e_2$] {};
\node (ar) at ( 1,1) [leaf,label=above:$r(e_2)$,label=below:$\scriptstyle \psi$] {};
\node (v) at ( 0,0) [sqvert] {};
\node (nl) at ( -1.2,-1) [leaf,label=below:$e_2$] {};
\node (nc) at ( 0,-1) [leaf,label=below:$e_2$] {};
\node (nr) at (1.2,-1) [leaf,label=below:$e_2$] {};
\draw [thick] (v) -- (al);
\draw [thick] (v) -- (ar);
\draw [thick] (v) -- (nl);
\draw [thick] (v) -- (nr);
\draw [thick] (v) -- (nc);
} \\
+
\frac{1}{3!2!} 
\mathpic{
[sqvert/.style={circle,draw=black,fill=white,thick,inner sep=0pt,minimum size=10pt},
rvert/.style={circle,draw=black,fill=black,thick,inner sep=0pt,minimum size=10pt},
leaf/.style={circle,draw=white,fill=white,inner sep=0pt,minimum size=0.1pt},
dott/.style={circle,draw=black,fill=black,inner sep=0pt,minimum size=1pt}
]
\node (al) at ( -1,1) [leaf,label=above:$r(e_2)$,label=below:$\scriptstyle \psi$] {};
\node (ar) at ( 1,1) [leaf,label=above:$r(e_2)$,label=below:$\scriptstyle \psi$] {};
\node (v) at ( 0,0) [sqvert] {};
\node (nl) at ( -1.2,-1) [leaf,label=below:$e_2$] {};
\node (nc) at ( 0,-1) [leaf,label=below:$e_2$] {};
\node (nr) at (1.2,-1) [leaf,label=below:$e_2$] {};
\draw [thick] (v) -- (al);
\draw [thick] (v) -- (ar);
\draw [thick] (v) -- (nl);
\draw [thick] (v) -- (nr);
\draw [thick] (v) -- (nc);
} +
\frac{1}{3!2!}
\mathpic{
[sqvert/.style={rectangle,draw=black,fill=white,thick,inner sep=0pt,minimum size=10pt},
rvert/.style={circle,draw=black,fill=white,thick,inner sep=0pt,minimum size=10pt},
leaf/.style={circle,draw=white,fill=white,inner sep=0pt,minimum size=0.1pt},
dott/.style={circle,draw=black,fill=black,inner sep=0pt,minimum size=1pt}
]
\node (au) at ( -1.7,1) [leaf,label=above:$e_2$] {};
\node (bu) at ( 1.7,1) [leaf,label=above:$e_2$] {};
\node (v1) at ( -0.8,0) [rvert,label={[label distance=-2pt]10:$e_2$}] {};
\node (v2) at ( 0.8,0) [rvert,label={[label distance=-2pt]170:$e_2$}] {};
\node (n2c) at (1.9,0) [leaf,label=right:$e_2$] {};
\node (ad) at ( -1.7,-1) [leaf,label=below:$e_2$] {};
\node (bd) at (1.7,-1) [leaf,label=below:$e_2$] {};
\draw [thick] (v1) -- (v2) node [below,align=center,midway] {$r^{22}$};
\draw [thick] (v1) -- (au);
\draw [thick] (v1) -- (ad);
\draw [thick] (v2) -- (bu);
\draw [thick] (v2) -- (bd);
\draw [thick] (v2) -- (n2c);
}\\
-\frac{1}{4}
\mathpic{
[sqvert/.style={rectangle,draw=black,fill=white,thick,inner sep=0pt,minimum size=10pt},
rvert/.style={circle,draw=black,fill=white,thick,inner sep=0pt,minimum size=10pt},
leaf/.style={circle,draw=white,fill=white,inner sep=0pt,minimum size=0.1pt},
dott/.style={circle,draw=black,fill=black,inner sep=0pt,minimum size=1pt}
]
\node (au) at ( -1.7,1) [leaf,label=above:$e_2$] {};
\node (bu) at ( 1.7,1) [leaf,label=above:$e_2$] {};
\node (v1) at ( -0.8,0) [rvert,label={[label distance=-2pt]10:$e_2$}] {};
\node (v2) at ( 0.8,0) [rvert,label={[label distance=-2pt]170:$e_2$}] {};
\node (n2c) at (1.9,0) [leaf,label=right:$r(e_2)$,label=above:$\scriptstyle \psi$] {};
\node (ad) at ( -1.7,-1) [leaf,label=below:$e_2$] {};
\node (bd) at (1.7,-1) [leaf,label=below:$e_2$] {};
\draw [thick] (v1) -- (v2) node [below,align=center,midway] {$r^{22}$};
\draw [thick] (v1) -- (au);
\draw [thick] (v1) -- (ad);
\draw [thick] (v2) -- (bu);
\draw [thick] (v2) -- (bd);
\draw [thick] (v2) -- (n2c);
}
- \frac{1}{4}
\mathpic{
[sqvert/.style={rectangle,draw=black,fill=white,thick,inner sep=0pt,minimum size=10pt},
rvert/.style={circle,draw=black,fill=white,thick,inner sep=0pt,minimum size=10pt},
leaf/.style={circle,draw=white,fill=white,inner sep=0pt,minimum size=0.1pt},
dott/.style={circle,draw=black,fill=black,inner sep=0pt,minimum size=1pt}
]
\node (au) at ( -1.7,1) [leaf,label=above:$e_2$] {};
\node (bu) at ( 1.7,1) [leaf,label=above:$e_2$] {};
\node (v1) at ( -0.8,0) [rvert,label={[label distance=-2pt]10:$e_2$}] {};
\node (v2) at ( 0.8,0) [rvert,label={[label distance=-2pt]170:$e_1$}] {};
\node (n2c) at (1.9,0) [leaf,label=right:$r(e_2)$,label=above:$\scriptstyle \psi$] {};
\node (ad) at ( -1.7,-1) [leaf,label=below:$e_2$] {};
\node (bd) at (1.7,-1) [leaf,label=below:$e_2$] {};
\draw [thick] (v1) -- (v2) node [below,align=center,midway] {$r^{21}$};
\draw [thick] (v1) -- (au);
\draw [thick] (v1) -- (ad);
\draw [thick] (v2) -- (bu);
\draw [thick] (v2) -- (bd);
\draw [thick] (v2) -- (n2c);
}\\
+ \frac{1}{8}
\mathpic{
[sqvert/.style={rectangle,draw=black,fill=white,thick,inner sep=0pt,minimum size=10pt},
rvert/.style={circle,draw=black,fill=white,thick,inner sep=0pt,minimum size=10pt},
leaf/.style={circle,draw=white,fill=white,inner sep=0pt,minimum size=0.1pt},
dott/.style={circle,draw=black,fill=black,inner sep=0pt,minimum size=1pt}
]
\node (au) at ( -1.5,1) [leaf,label=above:$e_2$] {};
\node (bu) at ( 3.1,1) [leaf,label=above:$e_2$] {};
\node (v1) at ( -0.8,0) [rvert,label={[label distance=-2pt]10:$e_2$}] {};
\node (v2) at ( 0.8,0) [rvert,label={[label distance=-2pt]170:$e_2$},label={[label distance=-2pt]10:$e_2$}] {};
\node (v3) at ( 2.4,0) [rvert,label={[label distance=-2pt]170:$e_2$}] {};
\node (n2u) at (0.8,1) [leaf,label=above:$e_2$] {};
\node (ad) at ( -1.5,-1) [leaf,label=below:$e_2$] {};
\node (bd) at (3.1,-1) [leaf,label=below:$e_2$] {};
\draw [thick] (v1) -- (v2)node [below,align=center,midway] {$r^{22}$};
\draw [thick] (v2) -- (v3)node [below,align=center,midway] {$r^{22}$};
\draw [thick] (v1) -- (au);
\draw [thick] (v1) -- (ad);
\draw [thick] (v3) -- (bu);
\draw [thick] (v3) -- (bd);
\draw [thick] (v2) -- (n2u);
}
+ \frac{1}{8}
\mathpic{
[sqvert/.style={rectangle,draw=black,fill=white,thick,inner sep=0pt,minimum size=10pt},
rvert/.style={circle,draw=black,fill=white,thick,inner sep=0pt,minimum size=10pt},
leaf/.style={circle,draw=white,fill=white,inner sep=0pt,minimum size=0.1pt},
dott/.style={circle,draw=black,fill=black,inner sep=0pt,minimum size=1pt}
]
\node (au) at ( -1.5,1) [leaf,label=above:$e_2$] {};
\node (bu) at ( 3.1,1) [leaf,label=above:$e_2$] {};
\node (v1) at ( -0.8,0) [rvert,label={[label distance=-2pt]10:$e_2$}] {};
\node (v2) at ( 0.8,0) [rvert,label={[label distance=-2pt]170:$e_1$},label={[label distance=-2pt]10:$e_1$}] {};
\node (v3) at ( 2.4,0) [rvert,label={[label distance=-2pt]170:$e_2$}] {};
\node (n2u) at (0.8,1) [leaf,label=above:$e_2$] {};
\node (ad) at ( -1.5,-1) [leaf,label=below:$e_2$] {};
\node (bd) at (3.1,-1) [leaf,label=below:$e_2$] {};
\draw [thick] (v1) -- (v2)node [below,align=center,midway] {$r^{21}$};
\draw [thick] (v2) -- (v3)node [below,align=center,midway] {$r^{12}$};
\draw [thick] (v1) -- (au);
\draw [thick] (v1) -- (ad);
\draw [thick] (v3) -- (bu);
\draw [thick] (v3) -- (bd);
\draw [thick] (v2) -- (n2u);
}
\end{multline*}
} 

Let us explain the notation. The coefficients in front of the graphs are just the inverse orders of the corresponding automorphism groups. The labels at the leaves are the ones coming from~$\mathcal{L}$, where we have left out the variables~$t^{d,\mu}$, and where we have replaced~$z$ by~$\psi$ to remind the reader that it keeps track of the power of $\psi$-class at that leaf. The decorations at the edges are split between the input, output and middle of the edge. For instance, an edge decorated by $r^{22} e_2 \otimes e_2$ is shown with a label~$e_2$ near the input and output of the edge, and a label~$r^{22}$ in the middle. The minus signs in the third line come from the minus sign in equation~\eqref{eq:E}. 

Note that in the original description of the algorithm, the first three graphs would have appeared as one graph with the sums of different decortions on the leaves, as would the second three graphs and also the last two graphs. We have used the linearity described in Remark~\ref{re:linear} to write them as the sums of graphs that appear above. 

To get the result of this computation we first note that $r^{\mu \nu} = r^{\mu}_\rho \eta^{\rho \nu}$. In our case this means that $r^{11} = 1$, and all other entries are~$0$. Thus, only the first three terms survive. Using either the dilaton equation or topological recursion, and using that $r(e_2) = e_1$ in this case, we see immediately that
$$
\tilde{\sigma}_5 = \sigma_5 + 10 \sigma_4 + 20 \sigma_3 . 
$$
This agrees with formula~\eqref{eq:inversion_H_contr} for the inversion transformed potential, as it should.

\subsection{Equivalence of descriptions}
It follows directly from the standard correspondence between differential operators of the type~\eqref{eq:Al-quintized} and Feyn\-man-type formulas in terms of graphs (\cite{Giv01b}, cf. \cite{PesSch95}) that the descriptions of the Givental group action given in Sections~\ref{sec:diff-oper} and~\ref{sec:expr-graphs} are equivalent.
For simplicity, we will first assume that $r_l (e_{\mathbf{1}}) = 0$ for all~$l$, allowing us to ignore the dilaton term. In that case, the only thing we have to show is that 
\begin{equation}\label{eq:lin-quadratic}
\begin{split}
\hat{R} := \exp \left( \sum_{d \geq 0, l \geq 1} t^{d,\nu} (r_l)^\mu_\nu \d_{d+l, \mu} + \frac{\hbar}{2} \sum_{i,j \geq 0} (-1)^{i+1} \d_{i,\mu} (r_{i+j+1})^{\mu \nu} \d_{j,\nu} \right)&\  \\
= \exp\left(\sum_{d \geq 0, l \geq 1} t^{d,\nu} (r_l)^\mu_\nu \d_{d+l, \mu}\right) \exp\left(\sum_{k,l \geq 0} (V_{k,l})^{\mu \nu} \d_{k, \mu} \d_{l,\nu} \right)&\ 
\end{split}
\end{equation}
where $\partial_{d,\mu}=\pdif{}{t^{d,\mu}}$ and $V_{k,l}$ is defined by
$$
-\frac{\hbar}{2} \frac{\exp(-r(-z))\exp(r(w)) - \Id}{z+w} = \sum V_{k,l} z^k w^l
$$
and we assume summation over repeated Greek indices (we will do so for the rest of this section). Equation~\eqref{eq:lin-quadratic} follows from the Campbell-Baker-Hausdorff formula in the following way. Write 
$$
X = \sum_{l\geq 1,d \geq 0} (r_l)^\mu_\nu t^{d,\nu} \d_{d+l,\mu}, \quad Y = \frac{\hbar}{2} \sum_{i,j \geq 0} (-1)^{i+1}  (r_{i+j+1})^{\mu \nu} \d_{i,\mu} \d_{j,\nu}  
$$
for the linear and quadratic parts in the exponent in~$\hat{R}$ respectively. Then~$Y$ commutes with any (iterated) commutator of $X$ and~$Y$ containing at least one copy of~$Y$. Therefore, it follows from Campbell-Baker-Hausdorff that $e^{X+Y} = e^X e^Z$, where 
\begin{multline}
Z := \frac{-e^{-\ad_X} + 1}{\ad_X} Y = \sum_{p \geq 0} \frac{(-1)^p (\ad_X)^p}{(p+1)!} Y \\
= \frac{\hbar}{2} \sum_{p \geq 0} \sum_{s + t = p} \sum_{f_1, \ldots, f_s \geq 0} \sum_{g_1, \ldots, g_t \geq 0} \sum_{i,j \geq 0}\frac{{\binom{p}{s}}}{(p+1)!} (-1)^{i + 1 + f_1 + \cdots + f_s + s} \cdot \\ 
\cdot (r_{f_s} \cdots r_{f_1} r_{i+j+1} r_{g_1} \cdots r_{g_t})^{\mu\nu}  \d_{i + f_1 + \cdots + f_s, \mu} \d_{j + g_1 + \cdots + g_t, \nu} .
\end{multline}
In the last equality we use the fact that~$r_l$ is symmetric when~$l$ is odd, and skew-symmetric when~$l$ is even. Writing $Z = \sum_{k,l} Z_{kl} \d_k \d_l$, it is easy to see that 
\begin{equation}
(z + w)\sum_{k,l} Z_{kl} z^k w^l = -\frac{\hbar}{2} (\exp(-r(-z))\exp(r(w)) - \Id )
\end{equation}
by expanding the right hand side and using the equality 
$$
\frac{1}{k! (n-k-1)! n} + \frac{1}{(k-1)!(n-k)! n} = \frac{1}{k! (n-k)!} . 
$$ 
This completes the proof of the equivalence of descriptions for the case where $r_l e_{\mathbf{1}} = 0$. 

For the general case, it is clear that replacing~$X$ by 
$$
\tilde{X} = X_1 + X_2 = - \sum_{l\geq 1} (r_l)^\mu_{\bf{1}} \d_{l+1,\mu} + \sum_{l\geq 1, d \geq 0} (r_l)^\mu_\nu t^{d,\nu} \d_{d+l,\mu} ,
$$
will not affect any of the arguments made above. That is, since~$X_1$ commutes with~$Y$, the same argument proves that $e^{\tilde{X} + Y} = e^{\tilde{X}}e^{Z}$. Therefore, it remains to show that
\begin{align}
\exp\left(- \sum_{l\geq 1} (r_l)^\mu_{\bf{1}} \d_{l+1,\mu} \right. &+ \left. \sum_{d \geq 0, l \geq 1} (r_l)^\mu_\nu t^{d,\nu} \d_{d+l,\mu}\right) \\ \nonumber
= \exp&\left(\sum_{d \geq 0, l \geq 1} (r_l)^\mu_\nu t^{d,\nu} \d_{d+l,\mu}\right)\left(\sum_{l\geq 1} (W_l)^\mu_{\bf{1}} \d_{l,\mu} \right) 
\end{align}
where~$W_l$ is defined by
$$
\sum_{l \geq 1} W_l z^l = (-z) \left(\exp\left(\sum_{l \geq 1} r_l z^l \right) - \Id \right) . 
$$
Since~$X_1$ commutes with any iterated commutator of $X_1$ and~$X_2$ including~$X_1$ at least once, we have $e^{X_1 + X_2} = e^{X_2} e^T$, where
\begin{multline}
T := \frac{-e^{-\ad_{X_2}} + 1}{\ad_{X_2}} X_1 \\
= - \sum_{p,l} \sum_{f_1,\ldots,f_p} \frac{1}{(p+1)!} (r_{f_p} \cdots r_{f_1} r_l)^\mu_{\bf{1}} \d_{f_1 + \cdots + f_p + l + 1, \mu} = \sum_{l\geq 1} (W_l)^\mu_{\bf{1}} \d_{l,\mu} .
\end{multline}
This completes the proof of the equivalence of the graphical and operator representation of Givental's theory.


\section{Inversion transformation}
\label{sec:inversion}

The so-called \textit{inversion transformation} is an important example of a transformation that gives a discrete symmetry of Frobenius structures. Namely, if one applies this transformation to any given Frobenius manifold, the resulting object is again a Frobenius manifold.

It turns out that in terms of the Givental group action one can express this transformation in a particularly nice way.

Let us recall the definition of the inversion transformation (\cite{Dub96}). Given a Frobenius manifold $M$ with flat coordinates $(t^1,\dots,t^n)$ and potential $F$, this transformation consists of the following change of coordinates:
\begin{align*}
\hat t^1 &= \half \frac{t_\sigma t^\sigma}{t^n},\\ \nonumber
\hat t^\alpha &= \frac{t^\alpha}{t^n} ~{\rm for ~}\alpha \neq 1,\, n,\\ \nonumber
\hat t^n &= -\frac{1}{t^n},
\end{align*}
together with the following change of the potential and the metric:
\begin{align*}
\hat F(\hat t) &=
(t^n)^{-2}\left[ F(t) - \half t^1 t_\sigma t^\sigma\right] = (\hat t^n)^2 F +\half \hat t^1 \hat t_\sigma \hat t^\sigma ,\\ \nonumber
\hat\eta_{\alpha\beta} &= \eta_{\alpha\beta}.
\end{align*}
We will also need the inverse of the inversion transformation:
\begin{align*}
 t^1 &= \half \frac{\htt_\sigma \htt^\sigma}{\htt^n},\\ \nonumber
 t^\alpha &= -\frac{\htt^\alpha}{\htt^n} ~{\rm for ~}\alpha \neq 1,\, n,\\ \nonumber
 t^n &= -\frac{1}{\htt^n}.
\end{align*}

We prove the following
\begin{theorem}\label{thm:inversion}
The inversion transformation is given by the Givental transformation $\Rh = \exp\br{\sum_{k\geq 1} \br{r_kz^k}\hat{\ }}$ with
\begin{align*}
 r_1 &= \br{
\begin{array}{cccc}
0 & \dots & 0 & 1 \\
0 & \dots & 0 & 0 \\
\vdots & \ddots & \vdots & \vdots \\
0 & \dots & 0 & 0 
\end{array}},\\ \nonumber
 r_k & = 0, \quad k>1.
\end{align*}

More precisely, if $\Fh(\htt)$ is the inversion transformation of $F(t)$, then the local expansion of $\Fh(\htt)$ at $\br{0,\dots,0,-1}$ is the same as the genus zero part without descendants of the $\Rh$-transformed potential of the cohomological field theory corresponding to the local expansion of $F(t)$ at $\br{0,\dots,0,1}$.
\end{theorem}

\begin{proof}[Proof of Theorem~\ref{thm:inversion}]

We are going to check that the coefficients of the local expansion of $\Fh(\htt)$ at $\br{0,\dots,0,-1}$ and 
the coefficients of the genus zero part without the descendants of the $\Rh$-transformed cohomological field theory potential corresponding to the local expansion of $F(t)$ at $\br{0,\dots,0,1}$ agree.

Let us determine the coefficients of $\Fh$. Recall that in flat coordinates, the metric is given by $\eta_{\alpha \beta} = \delta_{\alpha + \beta, n+1}$. Thus, the potential has the form
\begin{equation}
\label{eq:OrigPot}
F(t) = \half t^1\br{t^1t^n+\dots+t^nt^1} - \half t^1 t^1 t^n + H\br{t^2,\dots,t^n},
\end{equation}
for some function~$H$.

Note that we consider cohomological field theories as well as Frobenius potentials to be defined up to addition of any terms of order $2$ or lower in $t$'s, so we disregard such terms here and below.

Computing the inversion-transformed potential, we have
\begin{multline} \label{eq:FInv}
\Fh(\htt) = \half \htt^1\br{\htt^1\htt^n+\dots+\htt^n\htt^1} - \half \htt^1 \htt^1 \htt^n + \\
+ \dfrac{1}{8\htt^n}\br{\htt^2\htt^{n-1}+\dots+\htt^{n-1}\htt^2}^2
+ \htt^n \htt^n H\br{-\dfrac{\htt^2}{\htt^n},\dots,-\dfrac{\htt^{n-1}}{\htt^n},-\dfrac{1}{\htt^n}}.
\end{multline}

Recall the correlator notation for the coefficients of the potential and denote by
$$
\frac{1}{|\Auth((\al))|} \ba{\htau_0\br{\al_1}\dots\htau_0\br{\al_N}}^\mathrm{I}_H
$$
the coefficient of $\htt^{\al_1}\dots \htt^{\al_N}$  in the inversion-transformed potential coming from the last term of (\ref{eq:FInv}), and by
$$
\frac{1}{|\Auth((\al))|} \ba{\htau_0\br{\al_1}\dots\htau_0\br{\al_N}}^\mathrm{I}_Q
$$
the coefficient of $\htt^{\al_1}\dots \htt^{\al_N}$ in the inversion-transformed potential coming from the second-to-last term of~\eqref{eq:FInv}, where~$|\Auth((\al))|$ denotes the number of automorphisms of the collection of indices~$\al_i$.

We are interested in the local expansion near~$\br{0,\dots,0,-1}$, so we put $\htt^n = -1 + \ep$. Then, for the last term we have
\begin{multline}
\label{eq:inversion_H_contr}
(1-\ep)^2 H\br{\dfrac{\htt^2}{1-\ep},\dots,\dfrac{\htt^{n-1}}{1-\ep},\dfrac{1}{1-\ep}} = \\ 
= \mathop{\sum}_{N+p\geq 3} \ \mathop{\sum}_{2\leq\al_1\leq\dots\leq\al_N\leq n-1}\dfrac{\htt^{\al_1}\dots\htt^{\al_N}\ep^p\br{1-\ep}^{2-p-N}}{|\Auth((\al))| \; p!} \mathop{H_{\al_1\dots\al_N \underbrace{\scriptstyle n\dots n}}}_{\quad\qquad p}  \\ 
=\mathop{\sum}_{\substack{N+p\geq 3\\ k \geq 0}} \ \mathop{\sum}_{2\leq\al_1\leq\dots\leq\al_N\leq n-1}\dfrac{\binom{N+k+p-3}{k}}{|\Auth((\al))| \; p!}  \mathop{H_{\al_1\dots\al_N \underbrace{\scriptstyle n\dots n}}}_{\quad\qquad p} \htt^{\al_1}\dots\htt^{\al_N}\ep^{p+k},
\end{multline}
where $H$ with subscripts stands for the value of the respective multiple partial derivative of $H$ taken at~$(0,\dots,0,1)$. In terms of correlators this means that
\begin{multline} \label{eq:InvHCorr}
\phantom{privet}\ba{\htau_0(\al_1) \dots \htau_0(\al_N) \br{\htau_0(n)}^{q}}^\mathrm{I}_H \\ 
= \mathop{\sum}_{p+k=q} \dfrac{q!}{p!}\binom{N+k+p-3}{k}\mathop{H_{\al_1\dots\al_N \underbrace{\scriptstyle n\dots n}}}_{\quad\qquad p} \phantom{privet !}
\end{multline}
for  $2~\leq~\al_1,\dots,~\al_N~\leq~n~-~1$.

For the second-to-last term we have 
\begin{multline}
\dfrac{1}{8\htt^n}\br{\htt^2\htt^{n-1}+\dots+\htt^{n-1}\htt^2}^2 =\\ -\mathop{\sum}_{\substack{2\leq \al\leq \beta\leq n+1-\beta \leq n+1-\al \leq n-1\\k}} \dfrac{1}{|\Auth_2(\al,\beta)|}\ep^k \htt^{\al}\htt^{n+1-\al}\htt^{\beta}\htt^{n+1-\beta},
\end{multline}
where $|\Auth_2(\alpha,\beta)|$ is defined in the following way. Define  $\bar{\al} = n+1-\al$ for any $2 \leq \alpha \leq n-1$. Then $|\Auth_2(\al,\beta)|=1$ if all four numbers $\al, \beta, \bar{\al}$ and $\bar{\beta}$ are pairwise different, $|\Auth_2(\al,\beta)|=2$ if two of them are equal, but not equal to the other two, and $|\Auth_2(\al,\beta)|=8$ if all of them coincide.
In terms of correlators, this means that for $2 \leq \al, \beta, \bar{\al},  \bar{\beta} \leq n-1$
\begin{multline}\label{eq:inv_trans_Q}
\ba{\htau_0 \br{\al} \htau_0 \br{\bar{\al}} \htau_0 \br{\be} \htau_0 \br{\bar{\be}} \br{\htau_0 \br{n}}^k}^\mathrm{I}_Q \\
= - \dfrac{k!\; |\Auth((\alpha, \beta, \bar{\alpha}, \bar{\beta}))|}{|\Auth_2(\al,\beta)|},
\end{multline}
while $Q$-correlators of any other form vanish.

Since the Givental transformation acts trivially on cubic terms, the part containing $\htt^1$ obviously coincides with what is coming from the Givental transformation.

Let us describe the situation on the Givental side. The main point we are going to use is the very simple form of the matrices~$r_l$, where the only non-zero entry of~$r_1$ is  $(r_1)^1_n = 1$, and $r_l$ is identically zero for all other~$l$. Furthermore, we are interested only in decorated graphs where $t^{d,\mu}$ with $d>0$ do not enter the decorations, since we aim at recovering the Frobenius potential, which is the genus zero part without descendants. Thus we will write $t^\mu := t^{0,\mu}$ to simplify expressions.

Taking all this into account, by equation~\eqref{eq:leaf_decor} we have $\mathcal{L} = z e_1 t^n + \sum_{\mu=1}^n e_\mu t^\mu$ for the decoration of ordinary leaves, and no dilaton leaves since the expression (\ref{eq:dshift_decor}) vanishes entirely in our case. Furthermore, for the internal edges we have $\mathcal{E} = - e_1 \otimes e_1$ by equation~\eqref{formulaE}.

Let us find which decorated graphs will give a nonzero contribution. We see that $z$ always comes coupled to $e_1$, which allows us to use the dilaton equation (\ref{eq:DilatonEq}) to express all graphs with $z$ entering their decorations in terms of graphs without $z$ entering their decorations.

Since the contraction with the tensor associated with a vertex is a linear operation, we can represent a given graph as a sum of $2^k$ graphs, where $k$ is the number of the leaves, such that instead of the sum $z e_1 t^n + \sum_{\mu=1}^n e_\mu t^\mu$ on each leaf we will have either just $z e_1 t^n$ or just $\sum_{\mu=1}^n e_\mu t^\mu$. Then the dilaton equation implies that the contribution of each of these $2^k$ graphs is a multiple of the contribution of a graph resulting from removal of all $z e_1 t^n$ leaves from the given graph. All these resulting graphs then obviously do not have any $z$ entering into their decorations.

Let us then find which graphs with no $z$ in the decorations  after using the dilaton equation can give a non-zero contribution. The claim is that they are either single-vertex ones with any number of leaves, or trivalent ones with no more than two internal edges going out of each vertex. Recall that the tensors $\mathcal{V}[n]$ appearing on $n$-valent vertices are built from the coefficients of the $n$th homogeneous part of the original potential. Then the claim follows from the form of the decorations we have on the internal edges, namely $-e_1 \otimes e_1$, and the fact that $t^1$ enters the the original potential only in cubic terms. More precisely, if a vertex has an internal edge going from it, then, since the corresponding tensor $\mathcal{V}[n]$ gets contracted with $ - e_1 \otimes e_1$, $n$ should be equal to $3$ because only $\mathcal{V}[3]$ has non-zero components with one of the indices equal to $1$.

Furthermore, if there is only one internal edge going from a given vertex, then there are two leaves attached to this vertex decorated by $\sum_{\mu=1}^n e_\mu t^\mu$. Taking the linearity into account, we can represent the given decorated graph as a sum of $n^2$ graphs for which these two leaves are decorated by $e_\mu t^\mu$ and $e_\nu t^\nu$ for $\mu,\nu \in \bd{1, \ldots, n}$ respectively. From the form of the original potential~\eqref{eq:OrigPot} it follows that out of these graphs only the ones with $2\leq \mu \leq n-1$ and $\nu=n+1-\mu$ give a non-zero contribution.

By similar considerations a vertex with more than two internal edges attached to it will contribute zero, and on a vertex with precisely two internal edges attached to it only $e_n t^n$ survives as the decoration of the single leaf attached to it. 

With help of the linearity property we now totally expand all the decorated graphs we have after applying the Givental transformation. By the consideration above, we are left with the following sum, where each graph is of course multiplied by the inverse order of its automorphism group. First, there are all possible graphs with one vertex and any number of leaves decorated by~$e_\mu t^\mu$ for any $\mu$ and any number of leaves decorated by~$z e_1 t^n$. Second, there are all possible trivalent graphs with at least one internal edge in total and no more than two internal edges going from each vertex, with $ - e_1 \otimes e_1$ decorating internal edges, $e_n t^n$ decorating the single leaf attached to a vertex with two internal edges, and $e_\mu t^\mu$ and $e_{n-\mu+1} t^{n-\mu+1}$ for $2\leq \mu\leq n-1$ decorating two leaves attached to a vertex with only one internal edge going from it. Furthermore, all graphs obtained from the above trivalent ones by adding any number of leaves decorated by~$z e_1 t^n$ to 
any number of vertices are also included in the sum. One can find the graphical representation of all of these graphs below.

Thus we have described all relevant graphs giving the Givental-transformed potential. Now let us show that the Frobenius potential recovered from them precisely coincides with the inversion-transformed potential.
 


The contribution coming from the one-vertex graphs turns out to coincide with the last term of (\ref{eq:FInv}).
More precisely, if we denote the contribution of these one-vertex graphs to the coefficient of $\htt^{\al_1}\dots\htt^{\al_N}\ep^q$ in the Gi\-ven\-tal-trans\-formed potential by 
$$
\frac{1}{|\Auth((\alpha))|\; q!} \ba{\htau_0(\al_1)\dots\htau_0(\al_N)\br{\htau_0(n)}^{q}}^\mathrm{G}_H $$
we have
\begin{multline}\label{eq:giv_transf_H}
\frac{1}{|\Auth((\al))| \; q!} \ba{\htau_0(\al_1) \cdots \htau_0(\al_N)\br{\htau_0(n)}^{q}}^{\mathrm{G}}_H = \\ 
=\dfrac{1}{|\Auth((\al))| }\mathop{\sum}_{p+k=q}\dfrac{1}{k!\,p!} \ba{\tau_0(\al_1) \cdots \tau_0(\al_N)\br{\tau_0(n)}^{p}\br{\tau_1(1)}^{k}}
\end{multline}
(here on the right hand side the correlator corresponds to original, non-transformed potential).
Using the dilaton equation \eqref{eq:DilatonEq} we get 
{\allowdisplaybreaks
\begin{multline}\label{eq:after_dilaton}
\ba{\htau_0(\al_1) \cdots \htau_0(\al_N) \br{\htau_0(n)}^{q}}^{\mathrm{G}}_H \\ 
= q! \mathop{\sum}_{p+k=q} \dfrac{\br{N+k+p-3} \cdots \br{N+p-2}}{p!\;k!} \cdot \\
\cdot\ba{\tau_0(\al_1) \cdots \tau_0(\al_N)\br{\tau_0(n)}^{p}}\\ 
= \mathop{\sum}_{p+k=q}\dfrac{q!}{p!}\binom{N+k+p-3}{k}\ba{\tau_0(\al_1)\dots\tau_0(\al_N)\br{\tau_0(n)}^{p}} \\ 
= \mathop{\sum}_{p+k=q}\dfrac{q!}{p!}\binom{N+k+p-3}{k}\mathop{H_{\al_1\dots\al_N \underbrace{\scriptstyle n\dots n}}}_{\quad\qquad p},
\end{multline}
} 
which exactly coincides with $\ba{\htau_0(\al_1)\dots\htau_0(\al_N)\br{\htau_0(n)}^{q}}^{\mathrm{I}}_H$ on the inversion-transformed side~\eqref{eq:InvHCorr}.

In terms of graphs, equations (\ref{eq:giv_transf_H}) and (\ref{eq:after_dilaton}) can be expressed in the following way:~{\allowdisplaybreaks \begin{align}
&\ba{\htau_0(\al_1) \dots \htau_0(\al_N)\br{\htau_0(n)}^{q}}^{\mathrm{G}}_H =:
\mathpic{
[sqvert/.style={rectangle,draw=black,fill=white,thick,inner sep=0pt,minimum size=10pt},
rvert/.style={circle,draw=black,fill=black,thick,inner sep=0pt,minimum size=10pt},
leaf/.style={circle,draw=white,fill=white,inner sep=0pt,minimum size=0.1pt},
dott/.style={circle,draw=black,fill=black,inner sep=0pt,minimum size=1pt}
]
\node (al) at ( -1,1) [leaf,label=above:$e_{\alpha_1}$] {};
\node (ar) at ( 1,1) [leaf,label=above:$e_{\alpha_N}$] {};
\node at ( 0,1) [dott] {};
\node at ( -0.2,1) [dott] {};
\node at ( 0.2,1) [dott] {};
\node (v) at ( 0,0) [sqvert,label=center:$\scriptstyle H$] {};
\node (nl) at ( -1,-1) [leaf,label=below:$e_n$] {};
\node (nr) at (1,-1) [leaf,label=below:$e_n$] {};
\node at ( 0,-1) [dott] {};
\node at ( -0.2,-1) [dott] {};
\node at ( 0.2,-1) [dott] {};
\node at ( 0,-1.6) {$\underbrace{\hspace{65pt}}$};
\node at ( 0,-2) {$q$};
\draw [thick] (v) -- (al);
\draw [thick] (v) -- (ar);
\draw [thick] (v) -- (nl);
\draw [thick] (v) -- (nr);
} \\ \notag
&= \mathop{\sum}_{p+k=q}\binom{q}{k}
\mathpic{
[sqvert/.style={rectangle,draw=black,fill=white,thick,inner sep=0pt,minimum size=10pt},
rvert/.style={circle,draw=black,fill=white,thick,inner sep=0pt,minimum size=10pt},
leaf/.style={circle,draw=white,fill=white,inner sep=0pt,minimum size=0.1pt},
dott/.style={circle,draw=black,fill=black,inner sep=0pt,minimum size=1pt}
]
\node (al) at ( -1,1) [leaf,label=above:$e_{\alpha_1}$] {};
\node (ar) at ( 1,1) [leaf,label=above:$e_{\alpha_N}$] {};
\node at ( 0,1) [dott] {};
\node at ( -0.2,1) [dott] {};
\node at ( 0.2,1) [dott] {};
\node (v) at ( 0,0) [rvert] {};
\node (nl) at ( -1.5,-1) [leaf,label=below:$e_n$] {};
\node (nr) at (-0.2,-1) [leaf,label=below:$e_n$] {};
\node at ( -0.8,-1) [dott] {};
\node at ( -1,-1) [dott] {};
\node at ( -0.6,-1) [dott] {};
\node at ( -0.85,-1.6) {$\underbrace{\hspace{45pt}}$};
\node at ( -0.85,-2) {$p$};
\node (pr) at ( 1.5,-1) [leaf,label=below:$e_1$,label={[label distance=-4pt]85:$\scriptstyle\psi$}] {};
\node (pl) at (0.2,-1) [leaf,label=below:$e_1$,label={[label distance=-4pt]85:$\scriptstyle\psi$}] {};
\node at ( 0.8,-1) [dott] {};
\node at ( 1,-1) [dott] {};
\node at ( 0.6,-1) [dott] {};
\node at ( 0.85,-1.6) {$\underbrace{\hspace{43pt}}$};
\node at ( 0.85,-2) {$k$};
\draw [thick] (v) -- (al);
\draw [thick] (v) -- (ar);
\draw [thick] (v) -- (nl);
\draw [thick] (v) -- (nr);
\draw [thick] (v) -- (pl);
\draw [thick] (v) -- (pr);
}\\ \notag
&=\mathop{\sum}_{p+k=q} \dfrac{q!}{p!}\binom{N+k+p-3}{k}
\mathpic{
[sqvert/.style={rectangle,draw=black,fill=white,thick,inner sep=0pt,minimum size=10pt},
rvert/.style={circle,draw=black,fill=white,thick,inner sep=0pt,minimum size=10pt},
leaf/.style={circle,draw=white,fill=white,inner sep=0pt,minimum size=0.1pt},
dott/.style={circle,draw=black,fill=black,inner sep=0pt,minimum size=1pt}
]
\node (al) at ( -1,1) [leaf,label=above:$e_{\alpha_1}$] {};
\node (ar) at ( 1,1) [leaf,label=above:$e_{\alpha_N}$] {};
\node at ( 0,1) [dott] {};
\node at ( -0.2,1) [dott] {};
\node at ( 0.2,1) [dott] {};
\node (v) at ( 0,0) [rvert] {};
\node (nl) at ( -1,-1) [leaf,label=below:$e_n$] {};
\node (nr) at (1,-1) [leaf,label=below:$e_n$] {};
\node at ( 0,-1) [dott] {};
\node at ( -0.2,-1) [dott] {};
\node at ( 0.2,-1) [dott] {};
\node at ( 0,-1.6) {$\underbrace{\hspace{65pt}}$};
\node at ( 0,-2) {$p$};
\draw [thick] (v) -- (al);
\draw [thick] (v) -- (ar);
\draw [thick] (v) -- (nl);
\draw [thick] (v) -- (nr);}
\end{align}
}

Now we look at the graphs with a non-zero number of internal edges, which turn out to correspond precisely to the second-to-last term of (\ref{eq:FInv}). By the discussion above about graphs, they contribute only to correlators of the form 
$$
\ba{ \htau_0 \br{\al} \htau_0 \br{\bar{\al}} \htau_0 \br{\be} \htau_0 \br{\bar{\be}} \br{\htau_0 \br{n}}^k}_Q^\mathrm{G}
$$ 
(where the index $\mathrm{G}$ attached to correlator means that it corresponds to the Givental-transformed potential, and the index $Q$ means that we take just the part coming from graphs with at least one internal edge), and this contribution is the following (for $2\leq \al, \beta, \bar{\al},  \bar{\beta} \leq n-1$):
{\allowdisplaybreaks
\begin{multline}\label{eq:giv_transf_Q}
\frac{\ba{ \htau_0 \br{\al} \htau_0 \br{\bar{\al}} \htau_0 \br{\be} \htau_0 \br{\bar{\be}} \br{\htau_0\br{n}}^k}^\mathrm{G}_Q}{k!\; |\Auth((\al,\beta, \bar{\al}, \bar{\beta}))|} \\ 
=  \dfrac{1}{|\Auth_2(\al,\beta)|}\Bigg(
\mathop{\sum}_{m_1+m_2=k}\dfrac{1}{m_1!\,m_2!}\ba{\tau_0\br{\al}\tau_0\br{\bar{\al}} \br{\tau_1\br{1}}^{m_1}\tau_0\br{\mu}} \\ 
\br{r_1}^{\mu\nu} \ba{\tau_0\br{\nu}\tau_0 \br{\be}\tau\br{\bar{\be}}\br{\tau_1\br{1}}^{m_2}}\\ 
+ 
\mathop{\sum}_{
\begin{smallmatrix}
m_1+m_2+m_3 \\
=k-1                
\end{smallmatrix}
}\dfrac{1}{m_1!\,m_2!\,m_3!}
\ba{\tau_0\br{\al} \tau \br{\bar{\al}} \br{\tau_1\br{1}}^{m_1} \tau_0 \br{\mu_1}} \br{r_1}^{\mu_1\nu_1}\\
\ba{\tau_0\br{\nu_1}\tau_0\br{n} \br{\tau_1\br{1}}^{m_2}\tau_0\br{\mu_2}} \br{r_1}^{\mu_2\nu_2}  
\ba{\tau_0\br{\nu_2}\tau_0\br{\be}\tau\br{\bar{\be}}
\br{\tau_1\br{1}}^{m_2}}\\ + \cdots \\ 
+\ba{\tau_0\br{\al}\tau_0\br{\bar{\al}}\tau_0\br{\mu_1}} \br{r_1}^{\mu_1\nu_1}  
\ba{\tau_0\br{\nu_1}\tau_0\br{n}\tau_0\br{\mu_2}} \br{r_1}^{\mu_2\nu_2} \cdots \\ 
\cdots \ba{\tau_0\br{\nu_k}\tau_0\br{n}\tau_0\br{\mu_{k+1}}} \br{r_1}^{\mu_{k+1}\nu_{k+1}} 
\ba{\tau_0\br{\nu_{k+1}}\tau_0\br{\be}\tau\br{\bar{\be}}}\Bigg) .
\end{multline}
}
The contribution of each product of correlators on the right hand side is given by $(-1)^p\,m_1!\cdot\dots\cdot m_{p+1}!$, where $p+1$ is the number of correlators in the product. Thus, the result is equal to 
\begin{multline}
\ba{\htau_0\br{\al}\htau_0\br{\bar{\al}}\htau_0\br{\be} \htau_0\br{\bar{\be}}\br{\htau_0\br{n}}^k}^\mathrm{G}_Q\\ = \dfrac{k! \; |\Auth((\al,\beta,\bar{\al},\bar{\beta}))|}{|\Auth_2(\al,\beta)|}\mathop{\sum}_{p=1}^{k+1} (-1)^p \binom{k+1}{p} \\
= -\dfrac{k! \; |\Auth((\al,\beta,\bar{\al},\bar{\beta}))|}{|\Auth_2(\al,\beta)|},
\end{multline}
which coincides with what we have on the inversion-transformed side~(\ref{eq:inv_trans_Q}). This concludes the proof.
In terms of graphs formula (\ref{eq:giv_transf_Q}) takes the following form:
{\allowdisplaybreaks
\def\sp{1.6+0.5}
\def\sph{0.8+0.25}
\def\ds{0.2}
\begin{multline}
\ba{ \htau_0 \br{\al} \htau_0 \br{\bar{\al}} \htau_0\br{\be} \htau_0\br{\bar{\be}}\br{\htau_0\br{n}}^k}^\mathrm{G}_Q =: \mathpic{
[sqvert/.style={rectangle,draw=black,fill=white,thick,inner sep=0pt,minimum size=10pt},
rvert/.style={circle,draw=black,fill=white,thick,inner sep=0pt,minimum size=10pt},
leaf/.style={circle,draw=white,fill=white,inner sep=0pt,minimum size=0.1pt},
dott/.style={circle,draw=black,fill=black,inner sep=0pt,minimum size=1pt}
]
\node (au) at ( -1.5,1) [leaf,label=above:$e_{n+1-\alpha}$] {};
\node (bu) at ( 1.5,1) [leaf,label=above:$e_{n+1-\beta}$] {};
\node (v) at ( 0,0) [sqvert,label=center:$\scriptstyle Q$] {};
\node (nl) at ( -0.5,-1) [leaf,label=below:$e_n$] {};
\node (nr) at (0.5,-1) [leaf,label=below:$e_n$] {};
\node at ( 0,-1) [dott] {};
\node at ( -0.2,-1) [dott] {};
\node at ( 0.2,-1) [dott] {};
\node at ( 0,-1.6) {$\underbrace{\hspace{35pt}}$};
\node at ( 0,-2) {$k$};
\node (ad) at ( -1.5,-1) [leaf,label=below:$e_\alpha$] {};
\node (bd) at (1.5,-1) [leaf,label=below:$e_\beta$] {};
\draw [thick] (v) -- (au);
\draw [thick] (v) -- (ad);
\draw [thick] (v) -- (bu);
\draw [thick] (v) -- (bd);
\draw [thick] (v) -- (nl);
\draw [thick] (v) -- (nr);
}\\ 
= \dfrac{k!\; |\Auth((\alpha,\beta,\bar{\al},\bar{\beta}))|}{|\Auth_2(\al,\beta)|}\Bigg( 
-\mathop{\sum}_{m_1+m_2=k}\dfrac{1}{m_1!\,m_2!}\mathpic{
[sqvert/.style={rectangle,draw=black,fill=white,thick,inner sep=0pt,minimum size=10pt},
rvert/.style={circle,draw=black,fill=white,thick,inner sep=0pt,minimum size=10pt},
leaf/.style={circle,draw=white,fill=white,inner sep=0pt,minimum size=0.1pt},
dott/.style={circle,draw=black,fill=black,inner sep=0pt,minimum size=1pt}
]
\node (au) at ( -2,1) [leaf,label=above:$e_{n+1-\alpha}$] {};
\node (bu) at ( 2,1) [leaf,label=above:$e_{n+1-\beta}$] {};
\node (v1) at ( -0.8,0) [rvert,label={[label distance=-2pt]10:$e_1$}] {};
\node (v2) at ( 0.8,0) [rvert,label={[label distance=-2pt]170:$e_1$}] {};
\node (n1l) at ( -1.2,-1) [leaf,label=below:$e_1$,label={[label distance=-4pt]95:$\scriptstyle\psi$}] {};
\node (n1r) at (-0.5,-1) [leaf,label=below:$e_1$,label={[label distance=-4pt]85:$\scriptstyle\psi$}] {};
\node at ( -0.85,-1) [dott] {};
\node at ( -0.95,-1) [dott] {};
\node at ( -0.75,-1) [dott] {};
\node at ( -0.85,-1.6) {$\underbrace{\hspace{27pt}}$};
\node at ( -0.85,-2) {$m_1$};
\node (n2l) at ( 1.2,-1) [leaf,label=below:$e_1$,label={[label distance=-4pt]85:$\scriptstyle\psi$}] {};
\node (n2r) at (0.5,-1) [leaf,label=below:$e_1$,label={[label distance=-4pt]95:$\scriptstyle\psi$}] {};
\node at ( 0.85,-1) [dott] {};
\node at ( 0.95,-1) [dott] {};
\node at ( 0.75,-1) [dott] {};
\node at ( 0.85,-1.6) {$\underbrace{\hspace{27pt}}$};
\node at ( 0.85,-2) {$m_2$};
\node (ad) at ( -2,-1) [leaf,label=below:$e_\alpha$] {};
\node (bd) at (2,-1) [leaf,label=below:$e_\beta$] {};
\draw [thick] (v1) -- (v2);
\draw [thick] (v1) -- (au);
\draw [thick] (v1) -- (ad);
\draw [thick] (v2) -- (bu);
\draw [thick] (v2) -- (bd);
\draw [thick] (v1) -- (n1l);
\draw [thick] (v1) -- (n1r);
\draw [thick] (v2) -- (n2l);
\draw [thick] (v2) -- (n2r);
}
\\ 
+\mathop{\sum}_{m_1+m_2+m_3=k-1}\dfrac{1}{m_1!\,m_2!\,m_3!}\mathpic{
[sqvert/.style={rectangle,draw=black,fill=white,thick,inner sep=0pt,minimum size=10pt},
rvert/.style={circle,draw=black,fill=white,thick,inner sep=0pt,minimum size=10pt},
leaf/.style={circle,draw=white,fill=white,inner sep=0pt,minimum size=0.1pt},
dott/.style={circle,draw=black,fill=black,inner sep=0pt,minimum size=1pt}
]
\node (au) at ( -2,1) [leaf,label=above:$e_{n+1-\alpha}$] {};
\node (bu) at ( 3.6,1) [leaf,label=above:$e_{n+1-\beta}$] {};
\node (v1) at ( -0.8,0) [rvert,label={[label distance=-2pt]10:$e_1$}] {};
\node (v2) at ( 0.8,0) [rvert,label={[label distance=-2pt]170:$e_1$},label={[label distance=-2pt]10:$e_1$}] {};
\node (v3) at ( 2.4,0) [rvert,label={[label distance=-2pt]170:$e_1$}] {};
\node (n1l) at ( -1.2,-1) [leaf,label=below:$e_1$,label={[label distance=-4pt]95:$\scriptstyle\psi$}] {};
\node (n1r) at (-0.5,-1) [leaf,label=below:$e_1$,label={[label distance=-4pt]85:$\scriptstyle\psi$}] {};
\node at ( -0.85,-1) [dott] {};
\node at ( -0.95,-1) [dott] {};
\node at ( -0.75,-1) [dott] {};
\node at ( -0.85,-1.6) {$\underbrace{\hspace{27pt}}$};
\node at ( -0.85,-2) {$m_1$};
\node (n2l) at ( 1.15,-1) [leaf,label=below:$e_1$,label={[label distance=-4pt]85:$\scriptstyle\psi$}] {};
\node (n2r) at (0.45,-1) [leaf,label=below:$e_1$,label={[label distance=-4pt]95:$\scriptstyle\psi$}] {};
\node (n2u) at (0.8,1) [leaf,label=above:$e_n$] {};
\node at ( 0.8,-1) [dott] {};
\node at ( 0.9,-1) [dott] {};
\node at ( 0.7,-1) [dott] {};
\node at ( 0.8,-1.6) {$\underbrace{\hspace{27pt}}$};
\node at ( 0.8,-2) {$m_2$};
\node (n3l) at ( 2.8,-1) [leaf,label=below:$e_1$,label={[label distance=-4pt]85:$\scriptstyle\psi$}] {};
\node (n3r) at (2.1,-1) [leaf,label=below:$e_1$,label={[label distance=-4pt]95:$\scriptstyle\psi$}] {};
\node at ( 2.45,-1) [dott] {};
\node at ( 2.55,-1) [dott] {};
\node at ( 2.35,-1) [dott] {};
\node at ( 2.45,-1.6) {$\underbrace{\hspace{27pt}}$};
\node at ( 2.45,-2) {$m_3$};
\node (ad) at ( -2,-1) [leaf,label=below:$e_\alpha$] {};
\node (bd) at (3.6,-1) [leaf,label=below:$e_\beta$] {};
\draw [thick] (v1) -- (v2);
\draw [thick] (v2) -- (v3);
\draw [thick] (v1) -- (au);
\draw [thick] (v1) -- (ad);
\draw [thick] (v3) -- (bu);
\draw [thick] (v3) -- (bd);
\draw [thick] (v1) -- (n1l);
\draw [thick] (v1) -- (n1r);
\draw [thick] (v2) -- (n2l);
\draw [thick] (v2) -- (n2r);
\draw [thick] (v2) -- (n2u);
\draw [thick] (v3) -- (n3l);
\draw [thick] (v3) -- (n3r);
}
\\ 
\phantom{privet!}+\quad \mathbf{\cdots}\\ 
+(-1)^{k+1}
\mathpic{
[sqvert/.style={rectangle,draw=black,fill=white,thick,inner sep=0pt,minimum size=10pt},
rvert/.style={circle,draw=black,fill=white,thick,inner sep=0pt,minimum size=10pt},
leaf/.style={circle,draw=white,fill=white,inner sep=0pt,minimum size=0.1pt},
dott/.style={circle,draw=black,fill=black,inner sep=0pt,minimum size=1pt},
placeh/.style={circle,draw=white,fill=white,inner sep=0pt,minimum size=0.1pt}
]
\node (au) at ( -2,1) [leaf,label=above:$e_{n+1-\alpha}$] {};
\node (bu) at ( \sp+3.6,1) [leaf,label=above:$e_{n+1-\beta}$] {};
\node (v1) at ( -0.8,0) [rvert,label={[label distance=-2pt]10:$e_1$}] {};
\node (v2) at ( 0.8,0) [rvert,label={[label distance=-2pt]170:$e_1$},label={[label distance=-2pt]10:$e_1$}] {};
\node (v3) at ( \sp+2.4,0) [rvert,label={[label distance=-2pt]170:$e_1$}] {};
\node (n2u) at (0.8,1) [leaf,label=above:$e_n$] {};
\node (v25) at ( \sp+0.8,0) [rvert,label={[label distance=-2pt]170:$e_1$},label={[label distance=-2pt]10:$e_1$}] {};
\node (n25u) at (\sp+0.8,1) [leaf,label=above:$e_n$] {};
\node (ad) at ( -2,-1) [leaf,label=below:$e_\alpha$] {};
\node (bd) at (\sp+3.6,-1) [leaf,label=below:$e_\beta$] {};
\node at ( 0.8+\sph,0) [dott] {};
\node at ( 0.8+\sph-\ds,0) [dott] {};
\node at ( 0.8+\sph+\ds,0) [dott] {};
\node (v21) at ( 1.4,0) [placeh] {};
\node (v24) at (\sp+0.2,0) [placeh] {};
\node at ( 0.8+\sph,-0.6) {$\underbrace{\hspace{160pt}}$};
\node at ( 0.8+\sph,-1) {$k+2 \ \mathrm{vertices}$};
\draw [thick] (v1) -- (v2);
\draw [thick] (v2) -- (v21);
\draw [thick] (v24) -- (v25);
\draw [thick] (v25) -- (v3);
\draw [thick] (v1) -- (au);
\draw [thick] (v1) -- (ad);
\draw [thick] (v3) -- (bu);
\draw [thick] (v3) -- (bd);
\draw [thick] (v2) -- (n2u);
\draw [thick] (v25) -- (n25u);
}\Bigg) \\
= -\dfrac{k! \; \Auth(\al,\be,\bar{\al},\bar{\be})}{\#\Auth_2(\al,\beta)} .
\end{multline}
}
\end{proof}

\begin{remark} Note that in this Section we used neither semi-simplicity of the Frobenius structure nor the Euler vector field. The Givental group element that we obtained acts perfectly without any extra assumptions, except for the analyticity of $F(t)$ at point $(0,\dots,0,1)$. Even this last assumption is not necessary due to the following reasons. Since inversion transformation is singular at the origin, in Dubrovin's original formulation analyticity at some point other than the origin is implicitly assumed. This domain of analyticity can very well not include $(0,\dots,0,1)$. However, one can deal with this in Givental approach by considering not only the action of $\hat R$-operator, but also the action of $\hat \Psi$-operator \cite{Giv01}, which has a  simple form. This will make formulas a bit less nice, so for this reason we consider here only the case when   $F(t)$ is analytical at $(0,\dots,0,1)$. Going to the more general case with the help of $\hat\Psi$-operator is rather straightforward.
\end{remark}

\section{Relation to  Schlesinger transformations}

In the semi-simple case, the inversion transformation of Frobenius structures originates from a Schlesinger transformation of a special differential operator \cite{Dub96}:
\begin{equation}
\Lambda = \partial_z - U - \frac{1}{z} [\Gamma(u),U], 
\end{equation}
where $U$ is the diagonal matrix of canonical coordinates
\begin{equation}
U = \br{
\begin{array}{ccc}
u^1& & \\
   &\ddots & \\
   &       & u^n
\end{array}
}
\end{equation}
and $\Gamma$ is the Darboux-Egoroff matrix.

With the help of the results of \cite{BurSha10} and our known form of $\Rh$-matrix we are now able to reproduce the formula for the Schlesinger transformation for the rotation coefficients $\ga_{ij}$ from \cite{Dub96}:
\begin{align}\label{eq:DubTraRot}
\hat \ga_{ij} &= \ga_{ij} - A_{ij},\\ \nonumber
A_{ij} &= \dfrac{\sqrt{\partial_i t_1 \partial_j t_1}}{t_1} 
\end{align}
in Givental approach. We prove the following
\begin{proposition}\label{prop:schlesinger}
$\Rh$-transformation of rotation coefficients gives
\begin{equation}
\hat \ga_{ij} = \ga_{ij}-\dfrac{\sqrt{\partial_i t_1 \partial_j t_1}}{1+t_1}.
\end{equation}
\end{proposition}

\begin{proof}[Proof of Proposition~\ref{prop:schlesinger}]
Following \cite{BurSha10}, for the infinitesimal deformation of $\ga$ we have (we write all of the indices explicitly in order to get all of the instances of the metric correctly):
\begin{align}
 (r_1z)\hat{\ }\; \gamma^{ij} &= - \br{\Psi_0}^i_\al \br{r_1}^\al_\be \eta^{\be \ga} \br{\Psi_0}^j_\ga,
 \\ \notag
 (r_1z)\hat{\ }\; \br{\Psi_0}^i_\al &= \br{\Psi_1}^i_\be\br{r_1}^\be_\al - \br{\Psi_0}^i_\be \br{r_1}^\be_\ga\eta^{\ga\de}\br{\Psi_0}^j_\de\de_{jk}\br{\Psi_1}^k_\al,
 \\ \notag
 (r_1z)\hat{\ }\; \br{\Psi_1}^i_\al &= \br{\Psi_2}^i_\be\br{r_1}^\be_\al - \br{\Psi_0}^i_\be \br{r_1}^\be_\ga\eta^{\ga\de}\br{\Psi_0}^j_\de\de_{jk}\br{\Psi_2}^k_\al,
 \\ \notag
 &\mathop{\vdots}
\end{align}

Here by $\Psi_i$, $i=0,1,2,\dots$, we denote the twisted wave functions of the multi-KP hierarchy as in~\cite{BurSha10}.

Taking into account that $r_1$ has only one nonzero element, we see that this chain actually terminates in the sense that $\Psi_2$ never enters the expression for the total deformation of $\ga$, and also taking into account that \cite{Dub96,Leu01}
\begin{equation}
\su{k=1}{n} \br{\Psi_0}^k_1 \br{\Psi_1}^k_1 = t_1,
\end{equation}
we arrive at the following formula for transformed rotation coefficients:
\begin{align}\label{eq:GivTraRot}
\hat\ga^{ij} & = \ga^{ij} - \br{\Psi_0}^i_1\br{\Psi_0}^j_1\br{1-t_1+\br{t_1}^2-\br{t_1}^3+\dots} 
\\ \notag
& = \ga^{ij} - \dfrac{\sqrt{\partial_i t_1 \partial_j t_1}}{1+t_1}.
\end{align}
Here we should recall that in order to get our $\Rh$-matrix, we made a shift to the point $\br{0,\dots,0,1}$. Due to flat metric being anti-diagonal with unit components, we also have $t_1=t^n$. This means that the right hand side of (\ref{eq:GivTraRot}) actually coincides with that of (\ref{eq:DubTraRot}), which proves the claim.
\end{proof}

\section{Implications for integrable hierarchies}

The result of Section \ref{sec:inversion} allows us to explicitly obtain in\-ver\-sion-trans\-formed Hamiltonians of the principal hierarchy. We prove the following
\begin{proposition}\label{prop:hierachies}
Linear span of $\Rh$-transformed Hamiltonians of the principal hierarchy coincides with the linear span of in\-ver\-sion-trans\-formed Hamiltonians obtained in \cite{LiuXuZha10}.
\end{proposition}

\begin{proof}[Proof of Proposition~\ref{prop:hierachies}]
In order to prove this proposition, we use the results of \cite{BurPosSha10} for the deformation of $\Omega_{\al,p;\beta,q}$ under Givental transformation, where
\begin{equation}
\Omega_{\al,p;\beta,q} = \pdifd{F_0}{t^{\al,p}}{t^{\beta,q}},
\end{equation}
where $F_0$ is the total genus zero potential with descendants.

In the case of genus zero and for our $\Rh$-operator, for the infinitesimal deformation of Hamiltonians
\begin{equation}
\th_{\al,p} = \Omo{\al}{p},
\end{equation}
we have (following \cite{BurPosSha10}):
\begin{equation}\label{eq:def-formula-th}
(r_1z)\hat{\ }\; \th_{\al,p}
= U\; \th_{\al,p} + \delta^n_{\al} \th_{1,p+1},
\end{equation}
where operator $U$ is given by
\begin{equation}
U = -v^n-\half \su{\ga=1}{n}v^{\ga}v^{n+1-\ga} \pdif{}{v^1} + v^n \su{\ga=1}{n}v^\ga\pdif{}{v^\ga}.
\end{equation}
This infinitesimal deformation can be exponentiated to give the in\-ver\-sion-trans\-formed Hamiltonians:
\begin{equation}\label{eq:OurHamTransf}
\hat\th_{\al,p}\br{\hv} = \br{\exp\br{U\br{v}} \th_{\al,p}\br{v} + \delta^n_{\al} \exp\br{U\br{v}}\th_{1,p+1}\br{v}}\bigg|_{v=\hv}.
\end{equation}

Now we are able to compare our results with the ones of \cite{LiuXuZha10}, where the in\-ver\-sion-trans\-formed Hamiltonians are given in a bit less explicit form:
\begin{align}
&\hat{\theta}^{LXZ}_{1,0}(\hat{v})=-\frac{1}{v^n},\quad \hat{\theta}^{LXZ}_{1,p}(\hat{v})=-\frac{\theta_{n,p-1}(v)}{v^n},\quad  p\geq 1, \\ \nonumber
&\hat{\theta}^{LXZ}_{\al,p}(\hat{v})=\frac{\theta_{\al,p}(v)}{v^n},\quad
 2\leq \al\leq n-1,\ p\geq 0, \\
&\hat{\theta}^{LXZ}_{n,p}(\hat{v})=\frac{\theta_{1,p+1}(v)}{v^n},\quad p\geq 0,\nonumber
\end{align}

Applying the inverse inversion transformation, we get (for $2\leq \al \leq n-1$)
\begin{align}
& \hat{\theta}^{LXZ}_{\al,p}(\hat{v}) = -\hat v^n \th_{\al,p}\br{\frac{\su{i=1}{n}\hv^i\hv^{n+1-i}}{2\hv^n},-\frac{\hv^2}{\hv^n},\dots,-\frac{\hv^{n-1}}{\hv^n},-\frac{1}{\hv^n}} \\ \notag
 &= (1-\ep)\th_{\al,p}\br{\hv^1-\half\frac{\su{i=2}{n-1}\hv^i\hv^{n+1-i}}{1-\ep},\frac{\hv^2}{1-\ep},\dots,\frac{\hv^{n-1}}{1-\ep},\frac{1}{1-\ep}}
\end{align}
Now it's easy to see that the operator from (\ref{eq:OurHamTransf}) makes exactly this change of variables in the function $\th_{\al,p}$, which proves the coincidence of Hamiltonians $\th_{\al,p}$ for $2\leq \al \leq n-1$. In an analogous way, for $\al=1$ and $\al=n$ we see that our Hamiltonians do not coincide with the ones of \cite{LiuXuZha10} but are instead certain linear combinations of them, which is perfectly valid due to the fact that only the linear span of the collection of Hamiltonians is unambiguously defined.
\end{proof}

\begin{remark} In principle, the result of~\cite{BurPosSha10} gives also a deformation formula for the Hamiltonians of the full Dubrovin-Zhang hierarchy that is reduced to Equation~\eqref{eq:def-formula-th} in genus $0$. An advantage of Equation~\eqref{eq:def-formula-th} is that it is an ODE whose right hand side is \emph{linear} in Hamiltonians, and therefore we can immediately write a nice closed formula for its solution. In the general case the right hand side appears to be quadratic. This still allows to integrate the corresponding ODE formally, but the resulting formulas don't say much about the inverse-transformed Hamiltonians. The same is true for the tau-functions.
\end{remark}

\end{document}